\documentclass{jfm}
\usepackage{graphicx}
\usepackage{epstopdf, epsfig}
\usepackage{rotating}
\usepackage{lscape}

\shorttitle{Instability under periodic internal solitary waves}
\shortauthor{A. Posada-Bedoya, J. Olsthoorn and L. Boegman}

\title{Boundary layer instability beneath periodic internal solitary waves}

\author{Andres Posada-Bedoya\aff{1}
  \corresp{\email{21afpb@queensu.ca}},
  Jason Olsthoorn\aff{1}
 \and Leon Boegman\aff{1}}

\affiliation{\aff{1}Department of Civil Engineering, Queen's University,
Kingston, Ontario, ON K7L 3N6, Canada}

\begin{document}

\maketitle

\begin{abstract}
We investigated the stability of the bottom boundary layer (BBL) beneath periodic internal solitary waves (ISWs) of depression over a flat bottom through two-dimensional direct numerical simulations. We explored the effects of variation in wave Reynolds number $Re_{ISW}$ and wave period on the nature of the instability, and energy production in the separated BBL. The instability characteristics and rate of vortex shedding of the BBL were strongly dependent on $Re_{ISW}$. The BBL was laminar and convectively unstable at $Re_{ISW}$ 90 and 300, respectively. At $Re_{ISW}=300$, the convective wave packet was periodically amplified by each successive ISW, until vortex-shedding occurred. This implies noise-amplification behavior and suggests that the discrepancies in the critical $Re_{ISW}$, for vortex shedding between lab and different numerical simulations, are due to differences in background seed noise. Instability energy decreased under the front shoulder of the ISW, analogous to flow relaminarization under a favorable pressure gradient. At larger $Re_{ISW}=900$, the BBL was initially convectively unstable, and then the instability tracked with the ISW, characteristic of global instability, regardless of the ISW periodicity. The simulated initial convective instability at both $Re_{ISW}$ 300 and 900 is in agreement with local linear stability analysis which predicts that the instability group speed is always lower than the ISW celerity. Increased free-stream perturbations and larger $Re_{ISW}$ shift the location of vortex shedding (and enhanced bed shear stress) closer to the ISW trough, thereby potentially changing the location of maximum sediment resuspension from the ISW, in agreement with field observations at higher $Re_{ISW}$.
\end{abstract}

\section{Introduction}
Internal solitary waves (ISWs) are nonlinear waves of large amplitude, with a vertical structure that displaces isopycnals either purely downward (waves of depression) or purely upward (waves of elevation), and are a common feature of stratified lakes, estuaries and the coastal ocean \citep{helfrich2006,lamb2014,boegman2019}. These waves transport energy from their generation sites over long distances. As they approach the coast, they interact with the bottom boundary layer (BBL), which is the region of the water column where the ISWs are affected by the presence of the seafloor \citep{trowbridge2018}. The BBL extracts energy and momentum from the ambient flow, often in the form of turbulent eddies which provide a mechanism for dissipating energy and transporting mass, heat, and momentum vertically in the water column. These fluxes have implications for basin-scale energy budgets, circulation, and water quality. 

\begin{figure}
  \centerline{\includegraphics[width=0.8\textwidth]{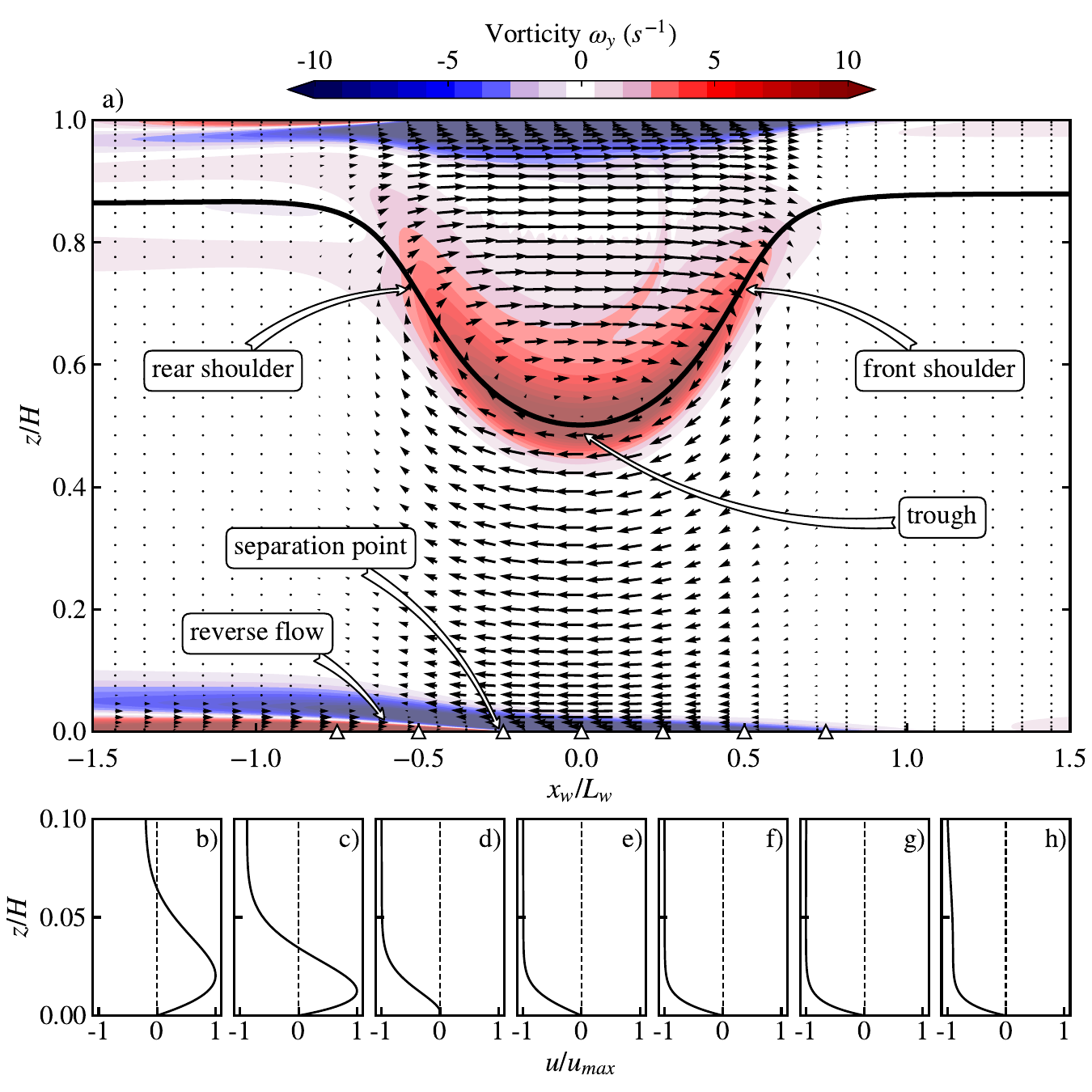}}
  \caption{(a) Vorticity field and velocity vectors of a typical rightward-propagating ISW. The black thick line indicates the pycnocline location. Triangles on the x-axis of (a) mark the location of the near-bed velocity profiles shown in panels (b-h) at $x/L_w$ -0.75, -0.5, -0.25, 0, 0.25, 0.5, and 0.75 respectively.}
\label{fig:intro}
\end{figure}

Although the BBL is characteristically forced by external flows (e.g., tides and surface waves), strongly nonlinear ISWs have the potential to impose unsteady and non-negligible horizontal pressure gradients that lead to rapid acceleration/deceleration of the flow  \citep{zulberti2020}. The ISW induces a horizontal current, which is maximum beneath the wave trough (Figure \ref{fig:intro}a). This streamwise velocity distribution imposes both favorable and adverse pressure gradients, on the BBL, under the front and rear shoulders of the ISW, respectively. If the adverse pressure gradient is sufficiently large, the boundary layer can separate (Figure \ref{fig:intro}b-d). Upstream of the separation point, the flow will reverse near the bed, forming a shear layer with an inflectional velocity profile (Figure \ref{fig:intro}b,c). This profile supports the amplification of small perturbations and is susceptible to exciting absolute instability \citep{huerre1990}. If the local absolute instability has a sufficient streamwise length scale, global instability can occur \citep{huerre1990}. The instability is said to be global, rather than local, due to the non-parallel streamwise dependence of the base flow. The signature of this global mechanism is the continuous excitation of self-sustained instabilities that grow and trigger vortex shedding that trails the ISW. Accompanying the passage of an ISW, localized sediment resuspension has been observed in the field \citep{bogucki1999,johnson2001,bogucki2005} and in the laboratory \citep{aghsaee2015,ghassemi2022} under the rear shoulder of the ISW. The resuspension has often been attributed to global instability of the separated region.

The production of unstable vortices by global instability of the separated BBL beneath ISWs was first suggested by \cite{bogucki1999} and then supported by numerical \citep[e.g.][]{diamessis2006,aghsaee2012,sakai2020} and laboratory \citep{carr2008} experiments. This is the commonly agreed upon instability mechanism resulting from the interaction of ISWs and the BBL \citep{boegman2019,zulberti2020}. However, as stated by \cite{boegman2019}, the possibility of global instability is somewhat surprising since the stratified shear instabilities, described in linear theory (i.e., by the Taylor–Goldstein equation), have propagation speeds that are much lower than the long wave speed. This, in turn, provides a strict lower bound on the ISW propagation speed. \cite{verschaeve2014} solved the Parabolized Stability equations for the linear instability of the spatially varying flow under an ISW. They concluded that the BBL under an ISW behaves as a noise amplifier, suggesting that the primary linear instability is convective. They showed steeper amplification of instabilities with increasing $Re$ and suggested that background seeding noise, in both lab and numerical domains, is critical to trigger flow stability. More recently, \cite{ellevold2023} conducted 2D Direct Numerical Simulation (DNS) of the BBL under ISWs of depression and attributed the instability to their numerical solver truncation error, suggesting a noise-amplifier behavior. However, as their analysis was non-modal, it remained a challenge to reconcile it with the modal stability concepts of convective, absolute and global stability. 

Similarly, lab-scale experiments and 2D DNS show discrepancies in predicting a critical Reynolds number for vortex shedding under ISWs. The experiments  \citep{carr2008,zahedi2021} agree that a critical momentum thickness Reynolds number $Re_{ISW}\approx200$ (defined in \ref{theory2}) is required for vortex shedding, which is much lower than the threshold proposed from 2D numerical simulations \citep{aghsaee2012}, which are also dependent on the pressure gradient $P_{ISW}$ (see Fig. 5b in \cite{zahedi2021}). More recently, 2D DNS by \cite{ellevold2023} showed good agreement predicting the threshold of instability of \cite{carr2008} experiments. \cite{ellevold2023} argued that the pycnocline thickness is an additional parameter relevant to the BBL stability (in addition to $Re_{ISW}$ and $P_{ISW}$) and suggested that the criterion proposed by \cite{aghsaee2012} is conservative. The reasons for these discrepancies, between experiments and different 2D DNS, remain unclear, but may be related to the effects of background seeding noise, as discussed above. 

More recently, the global instability paradigm was further challenged by novel high-resolution near-bed field measurements by \cite{zulberti2020} on the Australian continental shelf. They described sediment resuspension as a pumping mechanism resulting from the alternating compression and expansion of the highly turbulent BBL, as forced by the passage of an ISW packet. Under this pumping mechanism, there was no evidence of flow separation, Global Instability and sediment resuspension beneath the rear shoulder of the ISW \citep{aghsaee2015}. Rather, maximum near-bed sediment concentrations were observed beneath the ISW trough (where the flow was expected to relaminarize). A similar occurrence of near-bed sediment resuspension under the ISW trough was reported on the Portuguese shelf \citep{quaresma2007}. These different interpretations of flow instability and induced sediment resuspension, between low Reynolds number lab and high Reynolds number field studies, indicate that the nature of the BBL instability beneath an ISW, in response to changes in Reynolds number, requires further investigation. 

Overall, these discrepancies open questions regarding the nature of the instability and the relevance of comparing results obtained from numerical and experimental studies to each other and to field observations. The motivation for this study is to explore processes that could explain these discrepancies, which include: the strength of seed turbulence in the BBL \citep[e.g.][]{balzer2016,simoni2017}, $Re$ effects \citep[e.g.][]{balzer2016,simoni2017}, the presence of a background barotropic current \citep[e.g.][]{stastna2008,sakai2016}, relaminarization in the favorable pressure gradient beneath the front ISW shoulder \citep[e.g.][]{narasimha1979}, and wall roughness \citep{carr2010A,harnanan2017}. These processes remain under-investigated for the case of BBL stability beneath ISWs.  

We investigated BBL stability under periodic ISWs over a flat bottom for different $Re_{ISW}$ and wave periods with two objectives: (1) to investigate the effects of variation in $Re_{ISW}$ and ISW period on BBL stability, (2) to investigate the BBL stability of an ISW propagating into remnant seed turbulence from the wake of the preceding wave. The stability analysis of the BBL under periodic ISWs over a flat bottom is novel. Past numerical and lab studies have only considered laminar conditions preceding the passage of a lone ISW. We focused on the initial development of the instability, for which 2D DNS is suitable \citep[e.g.][]{diamessis2006,aghsaee2012} and invoked classical hydrodynamic stability theory \citep{drazin1981}. 

The paper is structured as follows: first, we present a brief review of instability theory concepts and nomenclature. We describe the problem and instability regimes over the parameter space investigated. We focused the description of the instability on individual cases representative of the relevant dynamics at each $Re_{ISW}$ and then provide context comparing across different $Re_{ISW}$ and wave periods. Finally, we discuss our results regarding the nature of the instability and potential implications for instability threshold definition and sediment transport. 

\section{Theoretical background}
\subsection{Linear convective, absolute and global instability} \label{theory1}
Here, we review the relevant theory on convective, absolute and global instability to provide context for our analysis. These concepts of local/global and absolute/convective instability provide the necessary theoretical framework to classify the flow stability depending on their qualitative behavior. Further detail can be found in \cite{huerre1990}, \cite{schmid2001} and \cite{chomaz2005}. 

The classical model for a steady parallel shear flow (independent of the streamwise direction) to transition from laminar to turbulence begins with an infinitesimal localized perturbation evolving over time and space. If the flow is unstable, the perturbation amplitude will grow over time. Moreover, a complete description requires determining the spatial-temporal evolution, which depends on the competition between the rate of advection of the instability energy, relative to the background flow, versus the rate of growth/amplification of the instability, often summarized in the impulse-response function. If the relative advection rate dominates, the instability may grow over time while being swept away from its generation site; this is a \emph{convective} instability. On the other hand, if the rate of growth dominates over the rate of advection, the instability will grow over time locally where it was introduced; this is an \emph{absolute} instability. 

Spatially varying flows are distinct from simple parallel flows, and have a streamwise variability in their velocity profile (like the BBL under ISW). For this case, unstable modes have a 2D spatial variability in the streamwise and vertical directions. These modes are termed \emph{global} because they have a coherent structure over a definite 2D region of the flow, at difference of a local instability which only refers to the instability of the local velocity profile. In this sense, globally unstable modes are the 2D equivalent of absolutely unstable modes in parallel shear flows. Moreover, both concepts are connected, as it has been established that a necessary condition for the existence of a global mode is the presence of a sufficiently large pocket of local absolute instability \citep{huerre1990}. 

In a convectively unstable flow, the evolution of the flow is highly sensitive to the amplitude and spectral content of external perturbations that are advected through, so they are also called \emph{noise amplifiers}. In this case, whether an instability grows to an observable level and triggers vortex shedding and/or transition to turbulence will depend on the initial amplitude of the perturbation, in turn depending on the initial level of noise. As environmental noise is often different across different experimental, numerical, and field settings, the definition of a general threshold for instability is challenging \citep[e.g.][]{verschaeve2014}. Rather, the absolute/global dynamics are intrinsic, spontaneous, and independent of the external noise \citep[e.g.][]{huerre1990}. Global modes act as self-excited wave-maker \emph{oscillators}, where instabilities are spontaneously and continuously generated, hence the evolution of vortices does not rely on the spatial amplification of external perturbations but rather on the growth of initial disturbances in time \citep{huerre1990}. It is, therefore, expected that global stability threshold parameterizations be equally valid in the lab and numerical domains. 

\subsection{Vortex shedding thresholds under ISWs}\label{theory2}
Using 2D DNS, \cite{aghsaee2012} argued that the BBL instability under an ISW is determined by the non-dimensional pressure gradient ($P_{ISW}$) and the momentum thickness Reynolds number ($Re_{ISW}$) at the separation point under the wave (Fig. \ref{fig:paramspace}): 

\begin{equation}
    P_{ISW}= \left ( U_2+c \right ) \frac{U_2}{L_w{g}'}
    \label{eq:PISW}
\end{equation}
\begin{equation}
    Re_{ISW}= U_2 \sqrt{\frac{L_w}{\nu\left ( U_2+c \right )}}
    \label{eq:ReISW}
\end{equation}
where $U_2$ is the absolute value of the maximum horizontal velocity at the wave trough, ${g}'=g\Delta\rho/\rho_0$ is the reduced gravity, $\nu$ is the kinematic viscosity, and $L_w$ is the horizontal wavelength scale \citep{michallet1999}:
\begin{equation}
    L_w=\frac{1}{a}\int_{-\infty}^{\infty}\eta_p(x)dx
    \label{eq:Lw}
\end{equation}
where $\eta_p(x)$ is the vertical displacement of the pycnocline. 
However, the observed critical $Re_{ISW}\approx200$ from lab experiments was much lower than the predicted by 2D DNS (Fig. \ref{fig:paramspace}). Typical values of $Re_{ISW}\gtrapprox 2500-3000$ (not shown) have been reported based upon field observations \citep{zulberti2020}. More recently, \cite{ellevold2023} reformulated \cite{aghsaee2012} criterion arguing that, in addition to $Re_{ISW}$ and $P_{ISW}$, the pycnocline thickness is also a relevant parameter for BBL stability. 

\section{Methods}
\subsection{Problem definition} 

\begin{figure}
  \centerline{\includegraphics[width=1\textwidth]{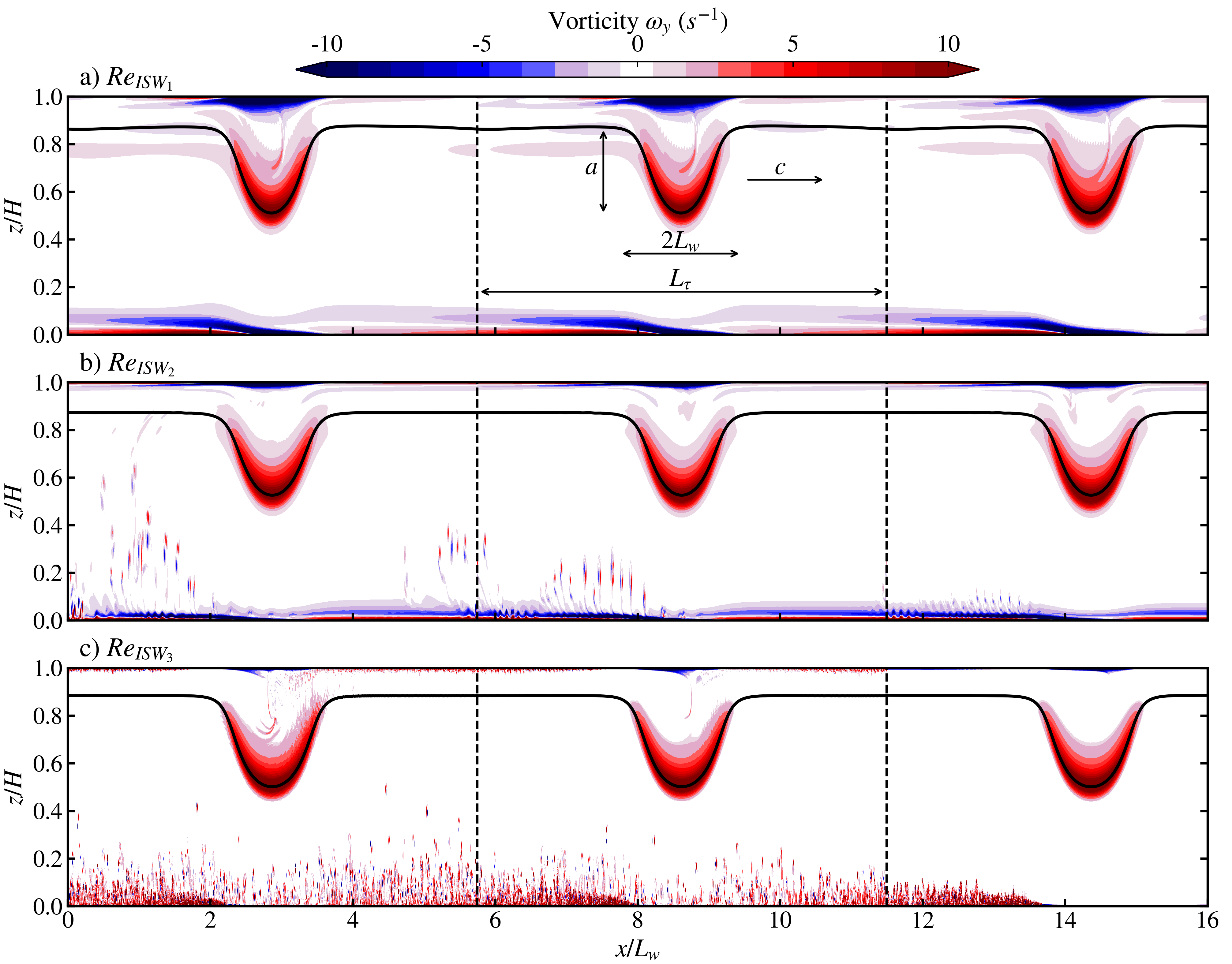}}
  \caption{Schematic of the research problem. Vorticity field of rightward propagating periodic ISWs at different $Re_{ISW}$ ($Re_{ISW_1} < Re_{ISW_2} < Re_{ISW_3}$). The black line represents the center of the pycnocline. Vertical dashed lines indicate the lateral limits of the periodic domain of length $L_{\tau}$. }
\label{fig:problem}
\end{figure}

We performed DNS of periodic ISWs of depression propagating over a flat bottom. We evaluated the effect of wave period, by modifying the length of our periodic domain ($L_{\tau}$), and Reynolds number ($Re_{ISW}$), by modifying the kinematic viscosity, on the BBL stability. A schematic of the problem is shown in Fig. \ref{fig:problem}, which illustrates the characteristic vortex shedding dynamics for the three different $Re_{ISW}$ regimes investigated here. Depending on the regime, subsequent periodic ISWs will encounter different characteristic trailing wakes from preceding ISWs. These will either be a stable laminar BBL at lower $Re_{ISW}$ (Fig. \ref{fig:problem}a), an unstable separated BBL with decaying vortices at intermediate $Re_{ISW}$ (Fig. \ref{fig:problem}b) or an energetic vortex wake at higher $Re_{ISW}$ (Fig. \ref{fig:problem}c). 

The periodic ISWs propagated along a quasi-two-layer density stratification defined via a hyperbolic tangent profile, widely used in numerical and laboratory studies \citep[e.g.][]{aghsaee2012}:
\begin{equation}
    \bar{\rho}(z)=\rho_0+\Delta\rho \tanh\left( \frac{z-z_{pyc}}{h_{pyc}}\right )
\end{equation}
where $\rho_0$ is a reference density (taken to be 1000 kg m$^{-3}$) and $\Delta\rho$ is the density jump across a pycnocline of thickness $2h_{pyc}$ centered at a depth $z_{pyc}$. Here, $z_{pyc}=4H/35$, $h_{pyc}=2H/35$, and $\Delta\rho/\rho_0$=120/1000, with $H$ being the total depth and $z$ increasing upwards from the bottom. 

The ISWs were initialized using a solution to the Dubreil-Jacotin-Long (DJL) equation:
\begin{equation}
    \nabla^2\eta+\frac{N^2(z-\eta)}{c^2}\eta =0
\end{equation}
\begin{equation}
    \eta=0\ at\ z=0,-H
\end{equation}
\begin{equation}
    \eta=0\ as\ x\rightarrow \pm\infty 
\end{equation}

\noindent where $\eta(x,z)$ is the vertical isopycnal displacement in the frame of reference of the wave, $c$ is the ISW phase speed, and $N$ is the Brunt-V\"ais\"al\"a frequency defined as
\begin{equation}
    N^2(z)=\frac{1}{\rho_0}\frac{d\bar{\rho}(z)}{dz}.
\end{equation}

The DJL equation was solved numerically using the algorithm of \cite{turkington1991} implemented by \cite{dunphy2011}. For this study, we chose a wave amplitude $a=0.37H$ with $L_w=2.5H$. The modeled periodic wave is a large amplitude ISW, similar to that of \cite{sakai2020}, and was selected to sustain its waveform over long propagation distances. 

\subsection{Parameter space}
\begin{figure}
\centerline{\includegraphics[width=0.8\textwidth]{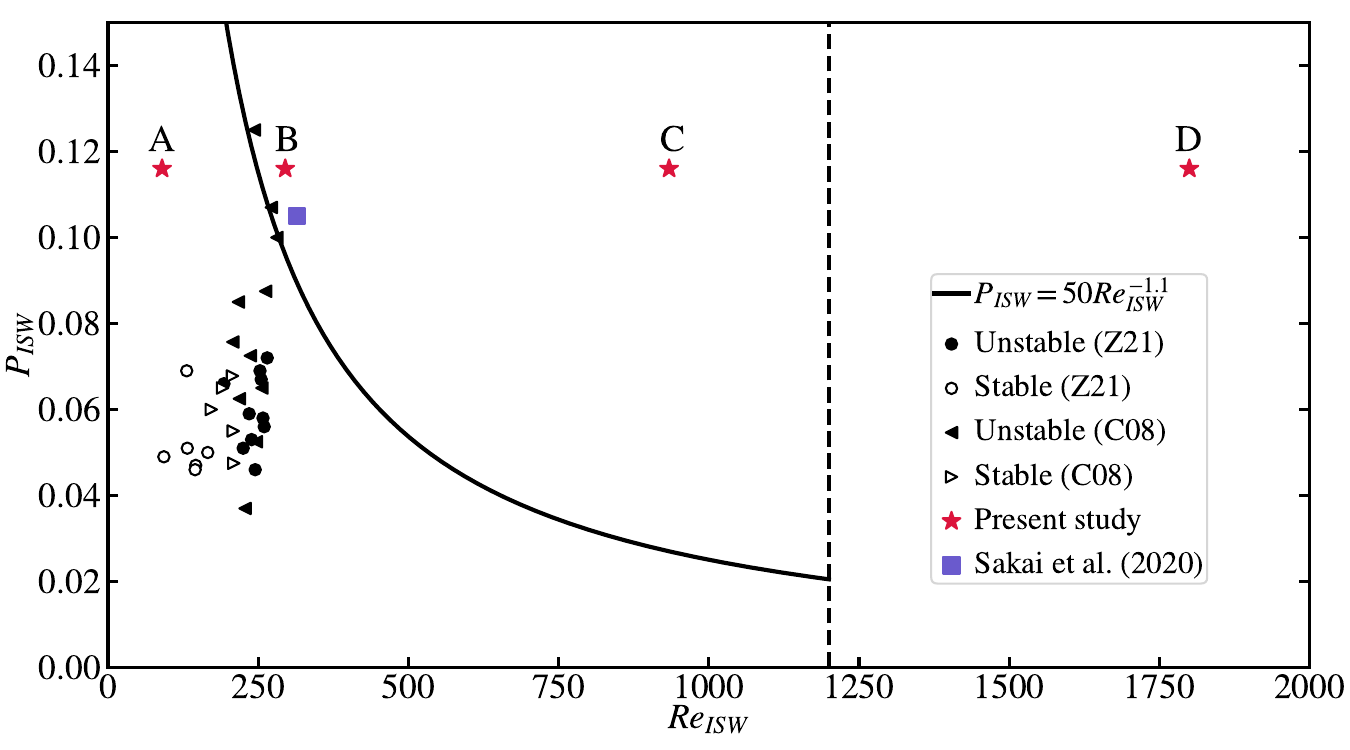}}
\caption{Simulated cases (red stars) plotted alongside the BBL stability curve proposed by \cite{aghsaee2012} as a function of $Re_{ISW}$ Vs $P_{ISW}$. We also include the ISW simulated by \cite{sakai2020} (blue square) and lab experiments reported by \cite{zahedi2021} (Z21) (circles) and \cite{carr2008} (C08) (triangles). The vertical dashed line denotes the estimated threshold of laminar-turbulent transition of the BBL according to \cite{aghsaee2012}.}
\label{fig:paramspace}
\end{figure}

We explored four $Re_{ISW}$ regimes based on the vortex-shedding threshold of \cite{aghsaee2012} (Fig. \ref{fig:paramspace}): (A) laminar $Re_{ISW}=90$, (B) marginally unstable, typical of lab-scale experiments $Re_{ISW}=300$, (C) highly unstable $Re_{ISW}=900$, and (D) higher than the laminar-turbulent transition for the boundary layer flow under ISWs ($Re_{ISW}\approx1200$) \citep{aghsaee2012} $Re_{ISW}=1800$. These Reynolds numbers were selected in order to cover a wide range, from a stable condition to a highly-unstable oceanographic regime. We modified the viscosity to set each target $Re_{ISW}$ (via eq. \ref{eq:ReISW}); all waves propagated on the same stratification and thus the same DJL wave properties, with constant $P_{ISW}=0.116$. 

For $Re_{ISW}$ A, B, and C, we simulated periodic domain lengths $L_\tau$, which varied between $L_\tau=3L_{w}$ and $L_\tau=12L_{w}$. The shortest ISW spacing was the limiting case and was just large enough, with respect to the ISW wavelength, to prevent any spurious interactions between the leading and trailing edges of each wave (ISW wavelength $\lambda_{ISW}\sim2L_w$). The largest spacing was four times the shortest case, large enough to demonstrate the effect of the periodicity, but limited by the required numerical resolution. For the largest $Re_{ISW}$ case (D), we only simulated a single wave with period $\tau$ to be a reference for the large $Re$ limit. Table \ref{table:1} summarizes the parameters for the cases simulated.

\begin{table}
  \begin{center}
\def~{\hphantom{0}}
  \begin{tabular}{lccccc}
      Case  & $Re_{ISW}$ & $L_{\tau}/L_w$ & $\nu$ (m$^2$/s) & Nx & Nz \\
      A1 & ~~90 & ~3 & $1\times10^{-5}$ & ~512 & ~256\\ 
      A2 & ~~90 & ~6 & $1\times10^{-5}$ & ~512 & ~256\\ 
      A3 & ~~90 & ~9 & $1\times10^{-5}$ & 1024 & ~256\\ 
      A4 & ~~90 & 12 & $1\times10^{-5}$ & 1024 & ~256\\
      B1 & ~300 & ~3 & $1\times10^{-6}$ & 1024 & ~512\\ 
      B2 & ~300 & ~6 & $1\times10^{-6}$ & 2048 & ~512\\ 
      B3 & ~300 & ~9 & $1\times10^{-6}$ & 2048 & ~512\\ 
      B4 & ~300 & 12 & $1\times10^{-6}$ & 4096 & ~512\\
      C1 & ~900 & ~3 & $1\times10^{-7}$ & 2048 & 1024\\ 
      C2 & ~900 & ~6 & $1\times10^{-7}$ & 4096 & 1024\\ 
      C3 & ~900 & ~9 & $1\times10^{-7}$ & 4096 & 1024\\ 
      C4 & ~900 & 12 & $1\times10^{-7}$ & 8192 & 1024\\
      D1 & 1800 & ~6 & $3\times10^{-8}$ & 8192 & 2048\\ 
  \end{tabular}
  \caption{Simulation parameters for each case. The domain length $L_\tau$ was a multiple of the wavelength scale $L_w$. The period of each ISW train was $\tau = L_\tau/c$.}
  \label{table:1}
  \end{center}
\end{table}

\subsection{Numerical simulations}
We numerically solved the two-dimensional incompressible Navier-Stokes equations under the Boussinesq approximation:
\begin{equation}
    \frac{\partial{u}}{\partial{t}} + u\frac{\partial{u}}{\partial{x}} + w\frac{\partial{u}}{\partial{z}}=-\frac{1}{\rho_0}\frac{\partial{p}}{\partial{x}}+\nu \nabla^2 u, 
    \label{eq:N-S1}
\end{equation}

\begin{equation}
    \frac{\partial{w}}{\partial{t}} + u\frac{\partial{w}}{\partial{x}} + w\frac{\partial{w}}{\partial{z}}=-\frac{1}{\rho_0}\frac{\partial{p}}{\partial{z}}+\nu \nabla^2 w-\frac{\rho g}{\rho_0},
    \label{eq:N-S2}
\end{equation}

\begin{equation}
    \frac{\partial{\rho}}{\partial{t}} + u\frac{\partial{\rho}}{\partial{x}} + w\frac{\partial{\rho}}{\partial{z}}=\kappa \nabla^2\rho,
    \label{eq:N-S3}
\end{equation}

\begin{equation}
    \frac{\partial{u}}{\partial{x}} + \frac{\partial{w}}{\partial{z}} = 0
    \label{eq:N-S4}
\end{equation}

\noindent where $(x,z)$ are the horizontal and vertical coordinates, $(u,w)$ are the associated velocity vectors, $t$ is time, $p$ is the pressure, $\rho$ is the fluid density, and $\kappa$ is the molecular diffusivity. For all cases, the simulated $Pr=\nu/\kappa=1$, however, we investigated various ratios between $\nu$ and $\kappa$ (between 1 and 10), and the results (not included) were insensitive to changes in $\kappa$ since the BBL is largely unstratified. 

The two-dimensional direct numerical simulations were conducted with the pseudospectral code SPINS \citep{subich2013}. Recent studies have shown the capability of SPINS to solve nonlinear internal waves problems in the laboratory scales to investigate wave-boundary interaction \citep{deepwell2021,hartharn-evans2022A} and boundary layer instability \citep{harnanan2017}. 

The computational domain was rectangular with depth $H$ and length $L_\tau$, the latter varied between cases to simulate different ISW periods. The horizontal domain was periodic to simulate the periodic passage of the ISWs. The initial condition was given by the DJL solution with the ISW in the middle of the domain. No-slip and no-flux boundary conditions were imposed on the top and bottom boundaries. A Chebyshev grid was employed in the vertical direction with a clustering of grid points near the top and bottom walls and a uniform grid was used in the horizontal direction. In all cases, initial noise of amplitude 0.001 was introduced to seed homogeneous perturbations in all cases. Grid resolutions ranged from 512 $\times$ 256 to 8192 $\times$ 2048 (Table \ref{table:1}). Grid-halving simulations verified grid independence at these resolutions. Time-dependent simulations were completed on the high-performance computing clusters of Compute Ontario. While we had originally planned three-dimensional simulations, it was computationally prohibitive to resolve Kolmogorov scales, in 3D, for our setups on the available computational resources. However, as we focused on the initial development of the instability, 2D simulations were sufficient to describe the essential dynamics of the primary instability and study its convective vs absolute nature prior to reaching a finite amplitude when it would trigger 3D secondary instabilities and transition to turbulence. \\

\subsection{Description of the BBL instability}

To describe the evolution of the instability, we separated base ($U$) and perturbation ($\hat{u}$) flow fields through low-pass and high-pass filtering of the instantaneous velocity field in wavenumber space. The cutoff wavenumber was determined from a wavelet analysis \citep{torrence1998}, which was also used to characterize the instability. We computed the evolution of wavelet spectra in the $k$-$x$ space of the near-bed vertically integrated horizontal velocity, which allowed us to track the position and wavenumber energy distribution of an unstable wave as it moves and grows over time. Here, we refer to the localized perturbation velocities, induced by the ISW-generated BBL instability, as an \emph{instability-generated wave packet}. By computing the wavelet spectra in $k$-$x$ space, we tracked wave energy packets to determine if the nature of the instability was convective or absolute. 

To understand the mechanisms for instability growth and its interaction with subsequent periodic ISWs, we computed the Reynolds-Orr energy budget \citep{schmid2001}:
\begin{equation}
    \frac{DE_v}{Dt}= -\hat{\mathcal{P}} - \hat{\varepsilon},
\label{eq:Rey-Orr1}
\end{equation}

\begin{equation}
    \hat{\mathcal{P}} = \int_{V}^{} \hat{u}_i \hat{u}_j\frac{\partial U_i}{\partial x_j} dV, 
\label{eq:Rey-Orr2}
\end{equation}

\begin{equation}
    \qquad \hat{\varepsilon} = \frac{1}{Re}\int_{V}^{}\left ( \frac{\partial \hat{u}_i}{\partial x_j} \right )^2 dV
\label{eq:Rey-Orr3}
\end{equation}

\noindent which describes the rate of change of the instability kinetic energy ($E_v=\frac{1}{2}\int_V\hat{u}_i\hat{u}_i dV$) due to its interaction with the base flow $U(x,z), W(x,z)$ ($-\hat{\mathcal{P}}$) and its viscous dissipation ($-\hat{\varepsilon}$) over the volume $V$. In two dimensions, $\hat{\mathcal{P}}$ is

\begin{equation}
    \hat{\mathcal{P}}=\int_{V}^{} \biggl( \underbrace{\hat{u} \hat{u}\frac{\partial U}{\partial x} }_{\mathcal{P}_{uu}} + \underbrace{\hat{w} \hat{w}\frac{\partial W}{\partial z}}_{\mathcal{P}_{ww}} +  \underbrace{\hat{u} \hat{w}\frac{\partial U}{\partial z} }_{\mathcal{P}_{uw}} + \underbrace{\hat{w} \hat{u}\frac{\partial W}{\partial x} }_{\mathcal{P}_{wu}} \biggr)dV
    \label{eq:prod1}
\end{equation}

As we are interested in the near-bed region, we computed these integrals over a sub-region of the domain. That is, we integrated over $L_{\tau}$, and vertically between $z=0$ and $z=h$, where $h$ was large enough to encompass the instability-generated wave packet, such that $E_v$ fluxes through the boundaries of $V$ were negligible. The separation of length scales between the ISW ($L_w$) and the instability-generated wave packet was large enough ($10^2-10^3$) such that visualized unstable oscillations ($\hat{u}$) and the budget (eq. \ref{eq:Rey-Orr1}) were not sensitive to the choice of filtering scales within $\approx$10-20\% above and below the cutoff value.  

\section{Results}

\subsection{Vorticity field over the parameter space}

\begin{landscape}
\begin{figure}
\centering
\includegraphics[width=1.53\textwidth]{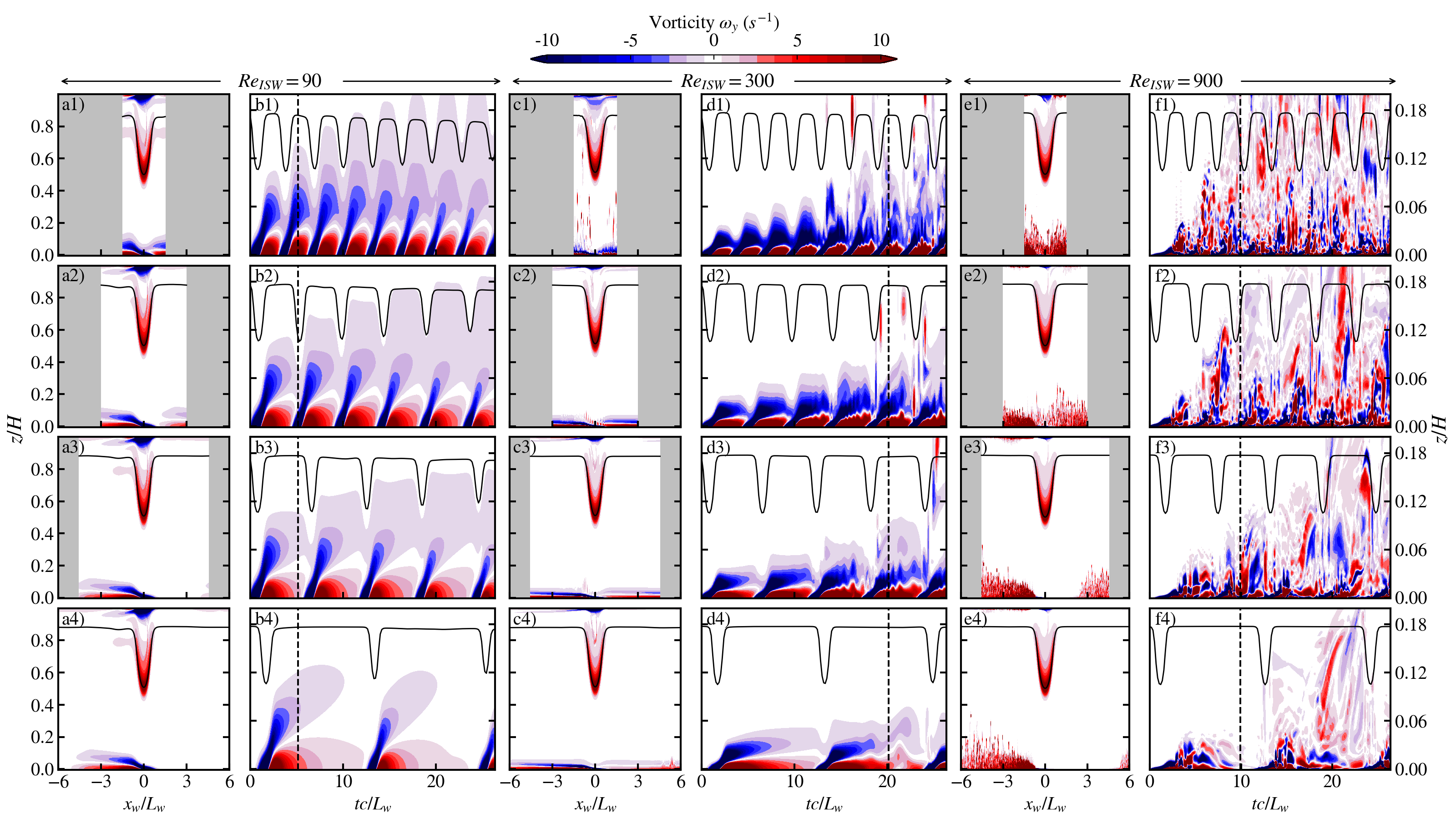}
\caption{(a,c,e) Snapshots and (b,d,f) near-bottom time-depth contours of the vorticity field for (a,b) $Re_{ISW}=90$, (c,d) $Re_{ISW}=300$ and (e,f) $Re_{ISW}=900$. Snapshots are in the wave reference frame, with $x_w=0$ corresponding to the ISW trough. The time of the snapshots (a,c,e) is indicated by a vertical dashed line in panels (b,d,f) respectively. Panels 1-4 correspond to increasing $L_{\tau}$ for each $Re_{ISW}$. In all panels, the continuous black line indicates the position of the pycnocline, whose scale reads on the left axis of panels (a). The scale of panels (b,d,f) reads on the right axis of panels (f). Corresponding movies of the vorticity field for simulations associated with panels (a1,b1), (a4,b4), (c1,d1), (c2,d2), (c3,d3), (c4,d4), (e1,f1), (e2,f2), (e3,f3), and (e4,f4) are provided in supplementary movies 1-10 respectively.}
\label{fig:vorty}
\end{figure}
\end{landscape}

The kinetic energy produced from the BBL depends upon both $Re_{ISW}$ and $L_{\tau}$. This can be illustrated with the vorticity field (Figure~\ref{fig:vorty}, supplementary movies 1-10). In all cases, the ISWs were stable and propagated rightward while remaining roughly unchanged over several wave periods $\tau$, except for a gradual reduction in wave amplitude due to friction. Behind each wave, the BBL separated due to the adverse pressure gradient under the wave. This formed two contiguous parallel vortex sheets of opposite sign (Fig.~\ref{fig:vorty}a1, at the bottom). Under the front shoulder of each propagating ISW, the flow accelerated leftwards leading to a compression of the boundary layer. Depending upon $Re_{ISW}$ and $L_{\tau}$, the BBL was laminar (panels a and b), had intermittent vortex shedding (panels c and d), or was continuously shedding vorticity (panels e and f). 

Within the parameter space evaluated here, there was a strong sensitivity of the boundary layer stability to $Re_{ISW}$, with the flow regime changing from stable to unstable between $Re_{ISW}$ 90 and 300. Further, the mild and intermittent vortex shedding at $Re_{ISW}$ 300 changed to energetic and continuous shedding at $Re_{ISW}$ 900. Higher $Re_{ISW}$ further increases the vortex shedding rate. 

The effect of $\tau$ was more subtle. The wave train period did not appear to have a direct effect on the stability of the BBL, but it did control the rate of energy production. This is particularly noticeable at $Re_{ISW}=300$. We discuss the three $Re_{ISW}$ cases in turn. 

The bottom boundary layer for the $Re_{ISW}=90$ cases was stable, with no signs of vortex shedding or unstable wave growth regardless of $L_{\tau}$ (Fig.~\ref{fig:vorty}a,b) (see supplementary movies 1 and 2). These cases had the largest effect of viscosity and exhibited the thickest boundary layer. As a result, these cases had the fastest frictional decrease in ISW amplitude. 

In each of the $Re_{ISW}=300$ cases, the BBL eventually became unstable, which resulted in vortex shedding (Fig.~\ref{fig:vorty}c,d)(see supplementary movies 3-6). As the region of instability lagged behind the ISWs, they periodically interacted with the instability, resulting in bursts of energy with the same periodicity as the ISWs. Vortices shed from the bed eventually encountered the pycnocline and distorted the ISW; the simulations were stopped at that time. Due to the periodic forcing, the vorticity of the shed vortices increased with ISW frequency, as can be seen by comparing the vorticity across panels c1-c4 in Fig.~\ref{fig:vorty}; vorticity decreased as $L_{\tau}$ increased. 

For the $Re_{ISW}=900$ cases (Fig.~\ref{fig:vorty}) vortex shedding tracked with the separated BBL under the ISWs (see supplementary movies 7-10). As a result, the near-bed region manifested continuous vortex shedding, independent of $L_{\tau}$. Vortex shedding was more vigorous and reached the pycnocline earlier than for the $Re_{ISW}=300$ cases with the same $L_{\tau}$. The growth rate of the instability was much larger than at $Re_{ISW}=300$, so the vortex shedding stage occurred much earlier regardless of $L_{\tau}$. \\

\subsection{Evolution of the BBL instability}
Motivated by the differences in vorticity production described above, we focus on the evolution of the BBL instability leading to vortex-shedding. The objective was to determine if the instability was convective or absolute/global. We considered two separate processes: (1) the unforced evolution of the instability after the first ISW in and before one ISW train period ($0<t/\tau<1$), and (2) the interaction of that instability with the following periodic ISWs ($t/\tau\geq1$). The instability excited by the first ISW was preceded by laminar conditions and was the same for all cases of a given $Re_{ISW}$ for a time $t/\tau<1$. This evolution was comparable to that in previous studies that considered the passage of a lone ISW of depression \citep{diamessis2006,aghsaee2012}. Conversely, the interaction of a trailing ISW with the wake generated by a leading wave has not been addressed before in the literature. We first describe the mechanics of the instability focusing on the largest $L_{\tau}$ case so we can track the instability over a long time before it interacts with the next ISW in the train and then compare all cases in the following section. We did not consider the $Re_{ISW}=90$ cases, as they did not exhibit boundary layer instability. 

\subsubsection{Individual wave}

\begin{figure}
\centering
\includegraphics[width=1\textwidth]{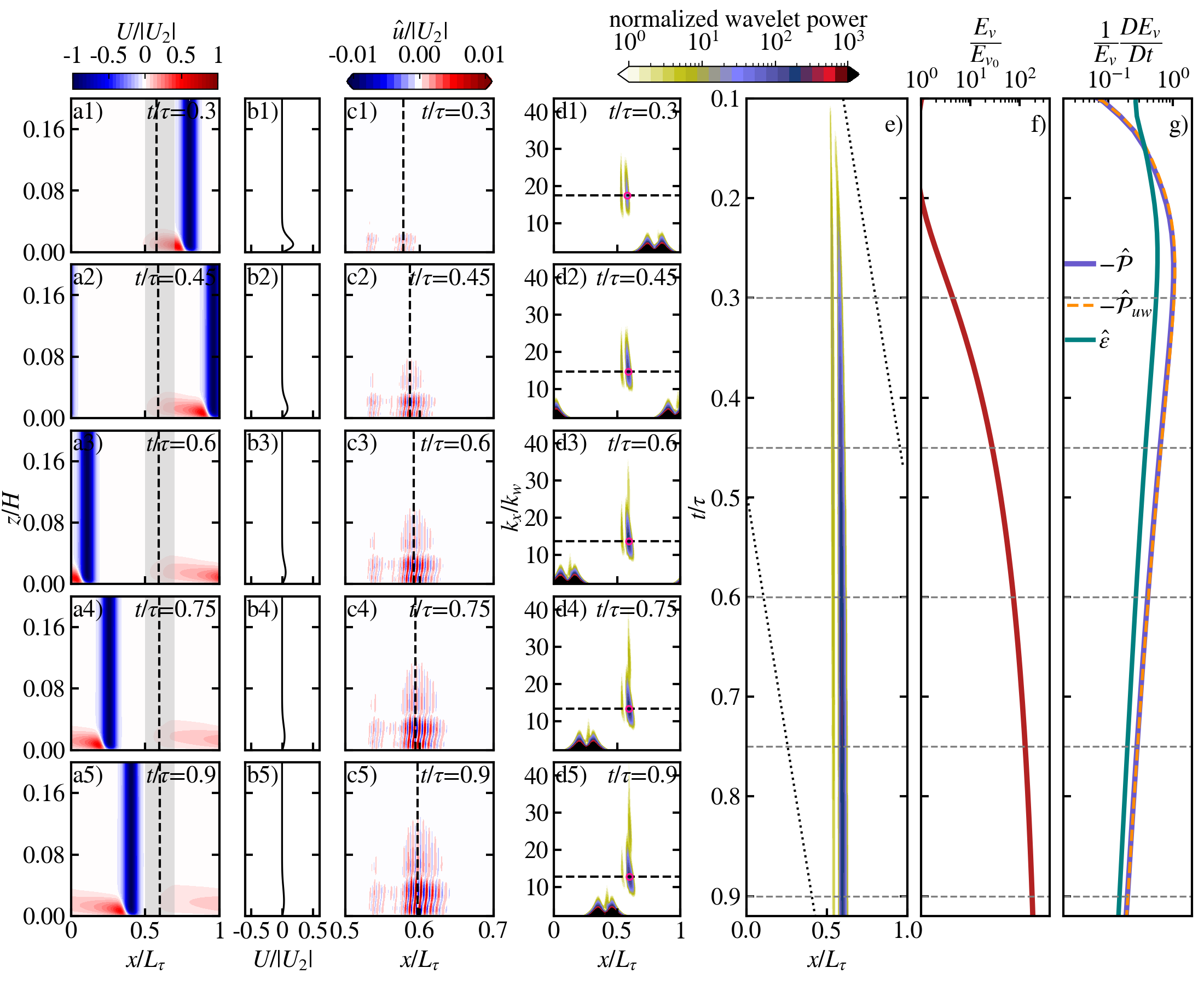}
\caption{Selected snapshots of near-bed (a) base ($U/\left | U_2 \right |$) and (c) instability ($\hat{u}/\left | U_2 \right |$) horizontal velocity field for case B4 (Table \ref{table:1}), $Re_{ISW}=300$. The time of the snapshot is indicated in each panel. Note the x-axis in panels (c) is different from elsewhere. Grey shadow in panels (a) indicates the observation area of panels (c). (b) $U/\left | U_2 \right |$ profiles at the location of maximum wavelet power, which is indicated with a vertical dashed line in (a) and (c). (d) Normalized wavelet spectra of the near-bed depth-integrated instantaneous horizontal velocity. The horizontal dashed line indicates the most energetic wavenumber and the circle, its position. (e) Hovm\"{o}ller plot of wavelet energy at the most energetic wavenumber component of the instability. The dotted line indicates the path of the ISW trough. Near-bed volume-integrated (f) instability kinetic energy ($E_v$) and the (g) relative rates of production ($\hat{\mathcal{P}}$) and dissipation ($\hat{\varepsilon}$). The corresponding movie of the instantaneous vorticity field for this simulation is provided in supplementary movie 6.}
\label{fig:instRe2-ind}
\end{figure}

In our $Re_{ISW}=300$ simulations (case B4, see Table \ref{table:1}), two distinct regions of velocity perturbation (instability-generated wave packets) were initially generated from the separated BBL under the rear shoulder of the ISW. While these packets initially moved slowly with the ISW, they were comparatively stationary and fell behind the ISW with a relative phase speed of $c_g/c\approx$0.04/0.30; persisting at $x/L_{\tau}\approx 0.6$. Figure~\ref{fig:instRe2-ind} shows the associated (a,b) base ($U$) and (c) perturbation ($\hat{u}$) velocity fields. Lagging behind the generating ISW suggests that the instability is convective, rather than the generally accepted global type.  

The wavelet analysis shows that the position of the instability-generated wave packet was nearly stationary (see Fig.~\ref{fig:instRe2-ind}d) and that the packet initially formed with a dominant wavenumber of $k_x/k_w\approx$18 at $t/\tau=0.3$ (Fig.~\ref{fig:instRe2-ind}d1) decreasing to $k_x/k_w\approx$13 at $t/\tau=0.9$ (Fig.~\ref{fig:instRe2-ind}d5), where $k_w=2\pi/L_w$. The wavelength of the most energetic mode was $\approx$ 20 times larger than $\Delta x$; therefore, the horizontal resolution was sufficient to resolve the instability. In the vertical direction, grid clustering near the wall was also sufficient to resolve the vertical structure. Moreover, the wave packet energy $E_v$ increased more than two orders of magnitude during the initial stage $0.2<t/\tau<0.8$ (Fig.~\ref{fig:instRe2-ind}e,f), with a maximum growth rate at $t/\tau\approx 0.25$.

The Reynolds-Orr budget for the instability shows that most of the energy growth resulted from the shearing term $\hat{\mathcal{P}}_{uw}=\hat{u}\hat{w}\partial U/\partial z$ (Fig.~\ref{fig:instRe2-ind}g). As the unstable wave packet fell behind the ISW, the base flow changed from being strongly sheared in the region of separation to a laminar ISW wake with gradually diminishing background shear (note the change velocity profiles in Fig.~\ref{fig:instRe2-ind}b1-b5). As a consequence, $-\hat{\mathcal{P}}-\hat{\varepsilon}$ reduced and the growth rate decreased (Fig. \ref{fig:instRe2-ind}g). The separated BBL never reattached to the bed (Fig.~\ref{fig:instRe2-ind}a1-a5); instead a reversed-flow shear layer decayed in the wake of the ISW. 

The main finding from the analysis of the $Re_{ISW}=300$ case, is that the instability-generated wave packet lagged behind the ISW. In the ISW reference frame, this wave packet was advected upstream from where it was generated. This behavior is characteristic of a \emph{convective instability}. \citep{huerre1990}. 

\begin{figure}
\centerline{\includegraphics[width=1\textwidth]{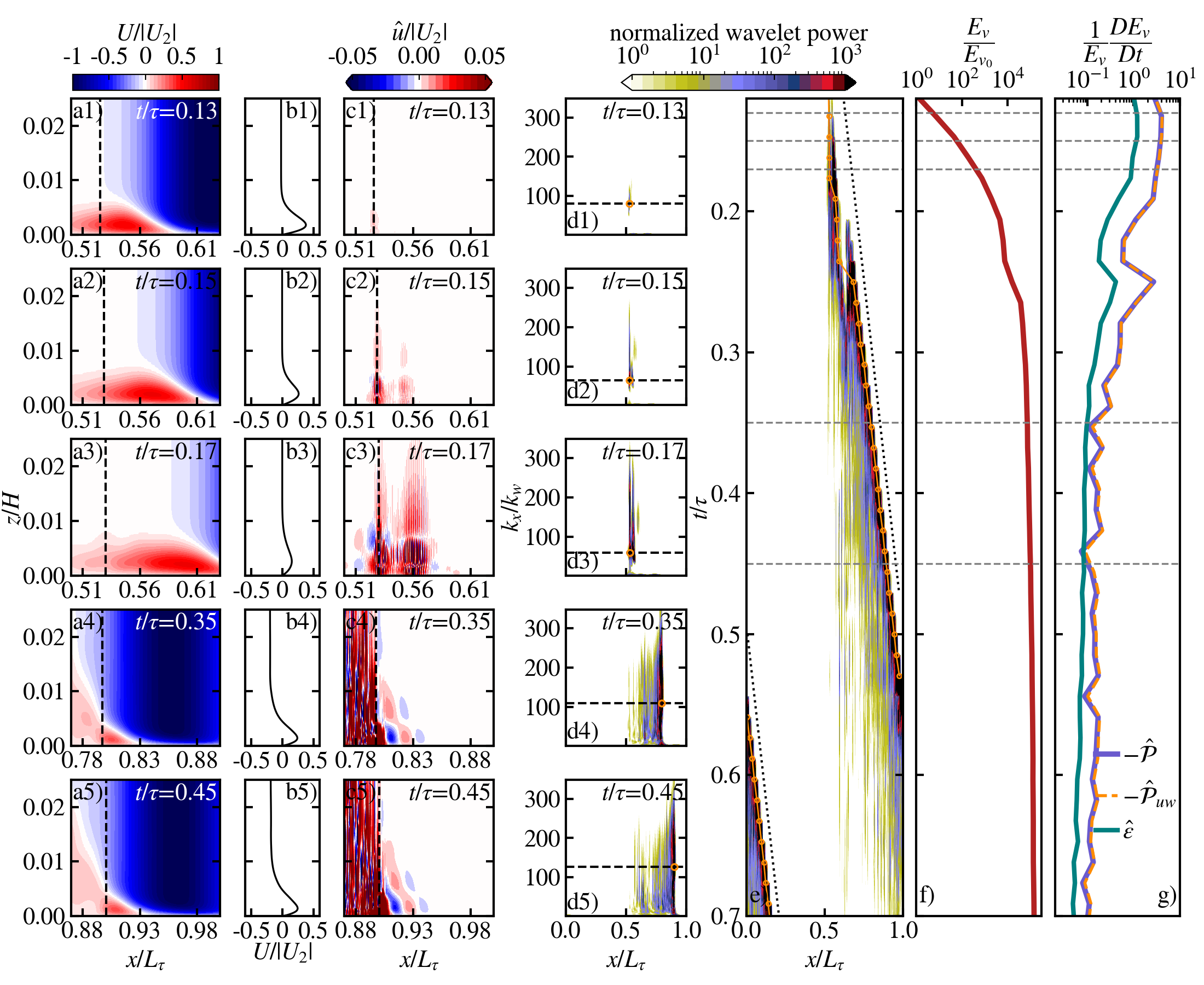}}
\caption{Same as in Fig. \ref{fig:instRe2-ind} for case C4 (Table \ref{table:1}), $Re_{ISW}=900$. In panel (e) the orange line indicates the location of the maximum wavelet power. Note that panels a1-a3 and c1-c3 show the same fixed region, whereas panels a4-a5 and c4-c5 show a region following the ISW. The corresponding movie of the instantaneous vorticity field for this simulation is provided in supplementary movie 10.}
\label{fig:instRe3-ind}
\end{figure}

For $Re_{ISW}=900$ (case C4, see Table \ref{table:1}), vortex shedding occurred much earlier than for $Re_{ISW}=300$. A close inspection of the early development of the flow shows two stages. First, as was the case for $Re_{ISW}=300$, two growing instability-generated wave packets emerged from the separation region, propagating at a much lower speed $c_g/c\approx$0.06/0.29, and falling behind the ISW at $x/L_{\tau}\approx0.53$ from $0.1<t/\tau<0.2$ (Fig.~\ref{fig:instRe3-ind}, panels 1-3). However, at $t/\tau\approx0.21$, a new region of instability formed at $x/L_{\tau}\approx0.65$ which then tracked with the ISW (Fig.~\ref{fig:instRe3-ind}, panels 4 and 5).

The initial instability had a broader wavenumber band ($k_x/k_w\approx$50-120) centered around a higher wavenumber ($k_x/k_w\approx$81) (Fig. \ref{fig:instRe3-ind}d1) than in the $Re_{ISW}=300$ case. The wider wavenumber spectrum can be identified in the different scales of periodic fluctuations composing the instability-generated wave packet (Fig. \ref{fig:instRe3-ind}c1-c3). As the initial instability fell behind, its dominant wavenumber decreased from $k_x/k_w\approx$81 at $t/\tau=0.13$ to $k_x/k_w\approx$61 at $t/\tau=0.17$, with energy spread over a wider wavenumber bandwidth $k_x/k_w\approx$30-300 (Fig. \ref{fig:instRe3-ind}d1-d3). After $t/\tau\approx0.2$, the second instability emerged and tracked with the ISW with a larger dominant wavenumber $k_x/k_w\approx$120, which remained roughly constant (Fig. \ref{fig:instRe3-ind}, panels 4 and 5). The background flow changed and the location of the maximum flow reversal shifted closer to the ISW trough (compare panels 1-3 versus 4-5 in Fig. \ref{fig:instRe3-ind}). As for the $Re_{ISW}$ 300 case, there was no BBL reattachment after separation. The wavelength of the most energetic mode was about 8 times larger than $\Delta x$, enough to resolve the instability. 

The kinetic energy of the initial instability grew more than three orders of magnitude from $0.1<t/\tau<0.2$ (Fig. \ref{fig:instRe3-ind}f), with a growth rate $\sim$5 times larger than for $Re_{ISW}=300$. The instability growth energy mostly came from the shearing term ($\hat{\mathcal{P}}_{uw}$) in the boundary layer wake flow of the ISW (Fig. \ref{fig:instRe3-ind}g). Therefore, as for the $Re_{ISW}=300$ case, the growth rate of the initial instability decreased as the background shear (Fig. \ref{fig:instRe3-ind}b1-b3) reduced whilst the instability initially fell behind. Between $0.2<t/\tau<0.25$ there was an increase in the growth rate coinciding with the emergence of the new region of instability tracking with the ISW. Around $t/\tau\approx 0.3$, the wave packet energy saturated, and continuous vortex shedding ensued, trailing the propagating ISW. 

The main finding from the analysis of the $Re_{ISW}=900$ case, is that two instabilities appeared: an initial instability in the BBL that was convectively unstable, and a secondary instability that continuously emanated from the separated BBL and tracked with the ISW. This secondary instability is characteristic of a \emph{global instability}. We comment further on these descriptions in the Discussion. 

\subsubsection{Periodic waves}\label{periodicwaves}
We now investigate how periodic ISWs interact with the instabilities, left behind in the wake of the preceding ISWs. For the $Re_{ISW}=300$ case, once the second periodic ISW reached the nearly-stationary instability-generated wave packet lagging behind the preceding ISW, different interactions occurred under the front and rear shoulders of the second ISW. 

\begin{figure}
\centerline{\includegraphics[width=1\textwidth]{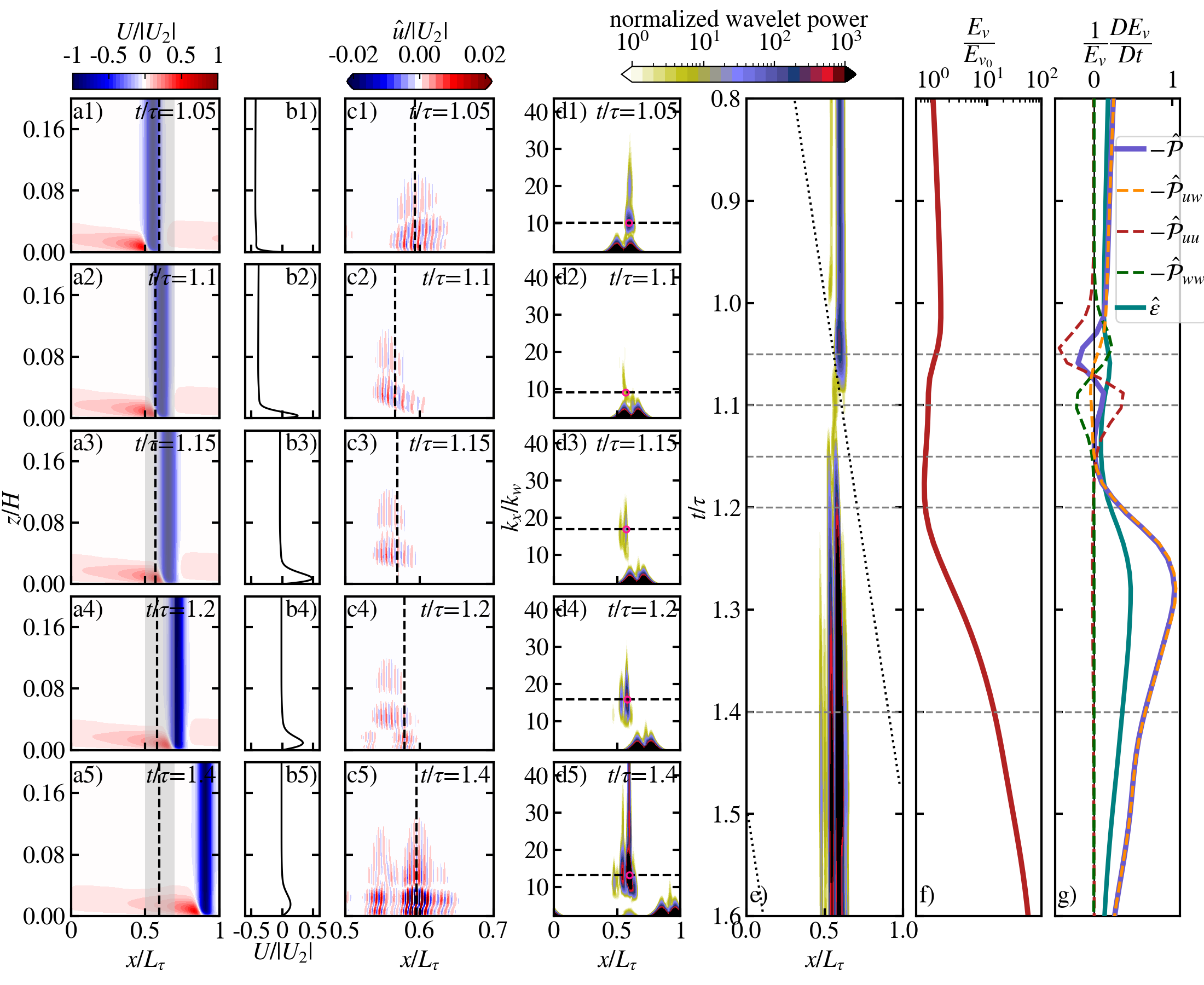}}
\caption{Same as in Fig. \ref{fig:instRe2-ind} for $0.8<t/\tau<1.6$. Case B4 (Table \ref{table:1}), $Re_{ISW}=300$. The corresponding movie of the instantaneous vorticity field for this simulation is provided in supplementary movie 6.}
\label{fig:instRe2-train}
\end{figure}

Under the front shoulder, of the rightward propagating ISW, the instability-generated wave packet experienced a leftward acceleration through the favorable pressure gradient, which forced it to stretch horizontally and squeeze vertically (Fig.~\ref{fig:instRe2-train}c1). The horizontal stretching led to a decrease of the most energetic wavenumber from $k_x/k_w\approx13$ (Fig. \ref{fig:instRe2-ind}d5) to $k_x/k_w\approx10$ (Fig. \ref{fig:instRe2-train}d1). The vertical squeezing pushed the instability closer to the bottom boundary, increasing viscous dissipation (see $\hat{\varepsilon}$ at $t/\tau\approx1.05$ in Fig.~\ref{fig:instRe2-train}g). The instability production term ($\hat{\mathcal{P}}$) was negative, dominated by the horizontal straining term ($-\hat{u}\hat{u}\partial U/\partial x<0$, $-\hat{\mathcal{P}}_{uu}$ in Fig.~\ref{fig:instRe2-train}g). This reduced the energy of the wave packet as it worked against the longitudinal straining induced by the base flow, acting in addition to viscous dissipation $-\hat{\varepsilon}<0$. The energy reduction, of the instability-generated wave packet, is analogous to the relaminarization experienced by a turbulent boundary layer under a favorable pressure gradient \citep{narasimha1979}. 

Under the rear shoulder of the ISW, the instability-generated wave packet decelerated through the adverse pressure gradient and was advected upwards. As the wave packet moved away from the wall, the rate of dissipation decreased and remained in a near balance with production, now positive. Once the entire wave packet was within the adverse pressure gradient, it seeded a new region of instability (see the near-bed region at 
$0.5<x/L_{\tau}<0.6$ and $z/H\approx0.02$ in panel c4 of Fig.~\ref{fig:instRe2-train}). The shear production ($-\hat{u}\hat{w}\partial U/\partial z>0$) grew exponentially, again becoming predominant in the budget as before the interaction. The new instabilities were superposed onto the initial instability-generated wave packet while their energy continued to grow (Fig.~\ref{fig:instRe2-train}, panel c5). In turn, each new periodic ISW triggered local velocity perturbations and amplified the local instability energy. This illustrates noise-amplifier behavior, which is characteristic of convective instability. At $t/\tau=1.4$, the dominant wavenumber $k_x/k_w\approx13$, as in the pre-interaction condition (Fig.~\ref{fig:instRe2-train}, panel d5). This was expected, considering that the base flow under the second ISW is very similar to that under the first wave, so the local unstable mode characteristics, including wavenumber, would also be roughly the same. 

For the $Re_{ISW}=900$ case, the peak velocity perturbation tracked with the ISW after $t/\tau\approx0.2$ (Fig. \ref{fig:instRe3-ind}). As such, subsequent periodic ISWs propagated over the decaying turbulent wake left behind by the first instability after vortex shedding, rather than the instability-generated wave packet itself, as we discuss in the following section.

\subsection{Free-stream perturbations and large Reynolds number effects}
\begin{figure}
\centerline{\includegraphics[width=0.9\textwidth]{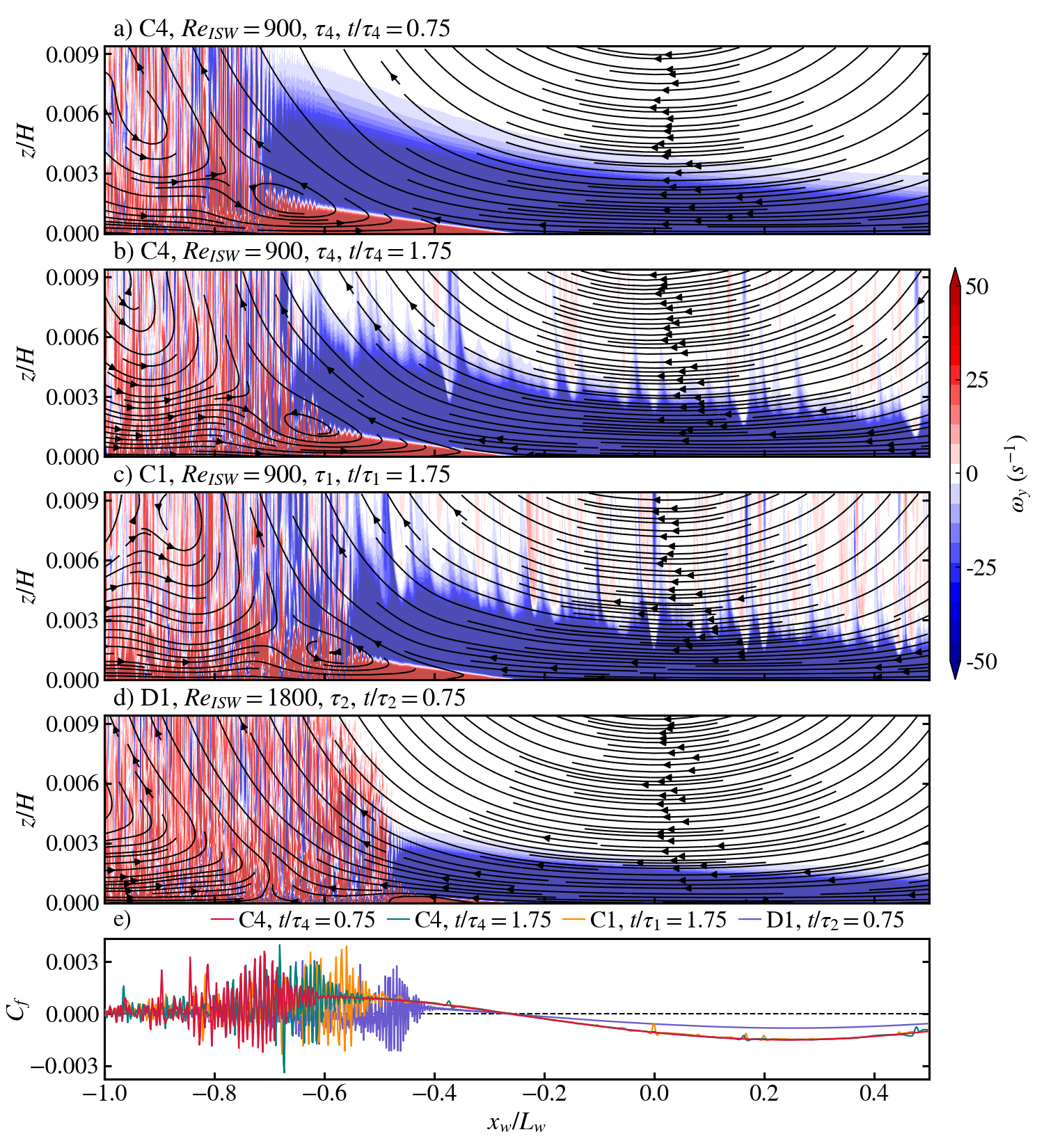}}
\caption{(a,b,c,d) Snapshots of the near-bed instantaneous vorticity field (contours) and base flow streamlines for case C4 ($Re_{ISW}=900$, $\tau_4$, largest train period) at (a) $t/\tau_4=0.75$ and (b) $t/\tau_4=1.75$, (c) case C1 ($Re_{ISW}=900$, $\tau_1$, shortest train period) at $t/\tau_1=1.75$, and (d) case D1 ($Re_{ISW}=1800$, $\tau_2$) at $t/\tau_2=0.75$. (e) Instantaneous bed shear stress coefficient $C_f=2\tau_b/\rho_0c^2$ along the streamwise axis associated with panels a-d. In all cases, the streamwise distance is scaled with the ISW length scale $L_w$ (eq. \ref{eq:Lw}) and abscissa $x_w/L_w=0$ corresponds to the ISW trough.}
\label{fig:noise}
\end{figure}

The occurrence of a second periodic ISW propagating over the decaying turbulent wake left behind the first ISW at $Re_{ISW}$ 900 provides an opportune scenario to investigate the influence of seed noise on BBL instability. We compared snapshots of the near-bed velocity, vorticity, and bed-stress fields under the ISW, at times both with and without the presence of a decaying wake, for the $Re_{ISW}=900$ scenarios: case C4 ($\tau_4$, largest wave period) at time $t/\tau=0.75$ without a wake (Fig.~\ref{fig:noise}a) and time $t/\tau=1.75$ with a wake (Fig.~\ref{fig:noise}b), and case C1 ($\tau_1$, shortest wave period) with a wake ($t/\tau=1.75$, Fig.~\ref{fig:noise}c). The shorter wave period case (C1) had a more energetic wake due to the shorter decay time before the passage of the next periodic ISW (cf. panels b and c in Fig.~\ref{fig:noise}). Here, we describe the bottom stress in terms of the instantaneous bed shear stress coefficient $C_f=2\tau_b/\rho_0c^2$, with $\tau_b$ being the bottom shear stress. 

The background flow in all cases had a reverse-flow vortex under the rear shoulder of the ISW, around the region where vortex shedding begins. For increasing levels of upstream perturbations, the reverse-flow vortex and the vortex-shedding location shifted closer to the ISW trough, which was more noticeable in the $C_f$ field (Fig.~\ref{fig:noise}e). Also, the reverse-flow vortex became smaller with its center closer to the bed with increasing wake energy (cf. panels a-c in Fig.~\ref{fig:noise}). We also compared the scenarios above with case D1 of $Re_{ISW}=1800$ at time $t/\tau=0.75$ without a wake (Fig.~\ref{fig:noise}d). The increase in Reynolds number produced a thinner boundary layer with the instability and vortex-shedding moving even closer to the ISW trough than in any $Re_{ISW}=900$ wake cases above. For $Re_{ISW}=1800$, the separated BBL reattached to the bed, forming a laminar separation bubble at $x/L_w\approx0.45$. 

In summary, the reverse-flow vortex became smaller and the vortex-shedding was closer to the ISW trough for higher $Re_{ISW}$ and increasing levels of seeding wake energy. Although this is a two-dimensional simulation, we would expect a similar response to external perturbations and high Reynolds number in three-dimensional (3D) flows, as supported by widely reported similar effects of increasing $Re$ and free-stream turbulence on the stability of laminar separation bubbles, both in experimental \citep{simoni2017} and 3D numerical \citep[e.g.][]{balzer2016} research. Moreover, while we believe the essential character of the instability is described by our 2D simulations, the reader should be aware that we are neglecting the fundamentally 3D processes associated with turbulent flows and dissipation. A complete description of the three-dimensional flow was computationally unfeasible given our available resources and is left for future work. 

\subsection{Amplification of the instability energy with $Re_{ISW}$ and $L_{\tau}$}

\begin{figure}
\centerline{\includegraphics[width=\textwidth]{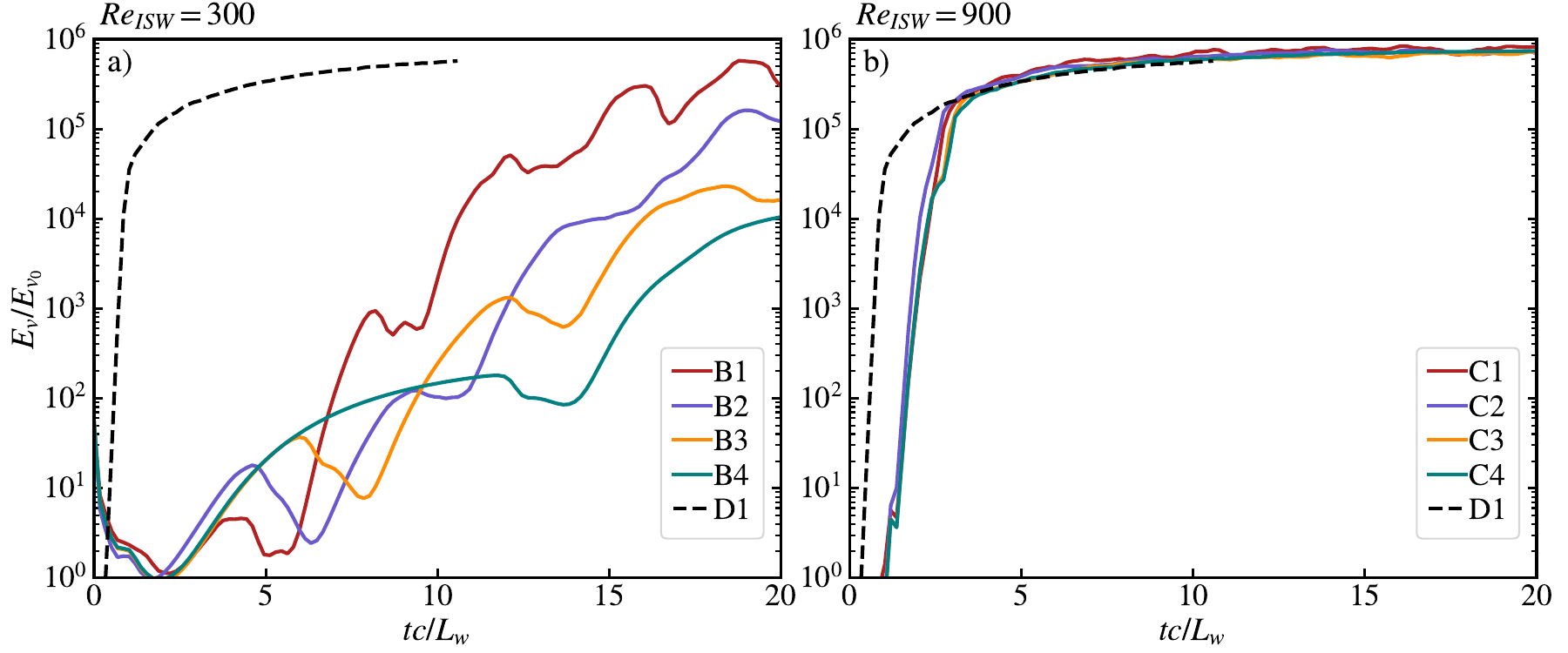}}
\caption{Evolution of normalized instability energy $E_v/E_{v_0}$ for (a) $Re_{ISW}=300$ and (b) $Re_{ISW}=900$. Cases are labeled according to Table \ref{table:1}. $E_{v_0}$ is constant and corresponds to the minimum $E_v$ of $Re_{ISW}=300$ cases. The black dashed line corresponds to the budget for $Re_{ISW}=1800$.}
\label{fig:budget-all}
\end{figure}

We now investigate how the instability kinetic energy $E_v$ (eq. \ref{eq:Rey-Orr1}) is periodically amplified by waves with different periods and Reynolds numbers (Fig. \ref{fig:budget-all}). We scaled $E_v$ by $E_{v_0}$, which is a constant that corresponds to the minimum $E_v$ of $Re_{ISW}=300$ cases. At $Re_{ISW}=300$, $E_v$ was reinforced with each periodic ISW (section \ref{periodicwaves}). As the trailing ISW approached a remnant instability-generated wave packet, $E_v$ initially decreased before amplifying from its initial value. After several wave periods, $E_v$ asymptoted to a finite saturation limit (Fig.~\ref{fig:budget-all}a). All $Re_{ISW}=300$ cases showed a similar evolution of $E_v$, with a periodic amplification of instability energy matching the ISW period. This indicated that the mechanistic description above can be extended to all $Re_{ISW}=300$ cases. In this regime, the instability was convective with a moderate growth rate, which was periodically amplified by each ISW in the train. Energy periodically built until it was large enough for the instability to trigger vortex shedding; thereafter, new bursts of vortex shedding occurred with each ISW passage. Because of this direct linkage between instability energy and ISW passage, the ISW periodicity is important in this flow regime. The role of the ISW period can be shown in the differences between the rates of accumulation of instability energy. The lowest wave period, with more frequent ISWs, reinforces energy into the instability more frequently and rapidly builds to the vortex-shedding stage, with higher enstrophy than in the largest train period case (Fig.~\ref{fig:budget-all}a).

For $Re_{ISW}=900$, vortex shedding tracked with the wave, and there was no apparent oscillatory behavior in the growth of $E_v$ (Fig.~\ref{fig:budget-all}b). Notably, this was true for all simulated ISW periods, with an almost identical evolution of $E_v$, also indicating that the mechanistic description presented for case C4 above can be extended to other cases with $Re_{ISW}=900$. The initial steep increase of $E_v$ converged towards an asymptotic final state with constant energy. The almost identical energy evolution suggests that the energy budget of the boundary layer instability is independent of the ISW train period. For $Re_{ISW}=1800$, $E_v$ also exhibited a steep initial increase followed by an asymptotic convergence to a final state of constant energy, which was reached faster than in the $Re_{ISW}=900$ cases. Interestingly, the final asymptotic saturation energy remains independent of $Re_{ISW}$, only differing in growth rate. It remains unclear to us why this saturation limit seems to be the same for all our $Re_{ISW}$. 

In summary, the effect of wave periodicity is only significant near a transition regime captured by the $Re_{ISW}=300$ cases. The instability triggered by an individual ISW is mild enough to not be shed as vortices before the next waves in the train interact with it and periodically reinforce it. Conversely, for a higher $Re$ regime, e.g. $Re_{ISW}=900$, the instability triggers vortex shedding before interacting with the following ISWs. The energy budget is nearly independent of the ISW period. 

\section{Discussion}

\subsection{Nature of the instability: convective or global/absolute}

We have shown the nature of the ISW-induced BBL instability to be dependent on the Reynolds number. Our numerical simulations clearly predicted laminar and convectively unstable regimes at $Re_{ISW}$ 90 and 300 respectively. At the larger $Re_{ISW}=900$, the BBL was initially convectively unstable ($t/\tau\lessapprox0.2$) (Figure \ref{fig:instRe3-ind}, panels 1-3), and then ($t/\tau\gtrapprox0.2$) a secondary instability arose that continuously tracked with the ISW, which is a feature that is characteristic of a global instability (Figure \ref{fig:instRe3-ind}, panels 4-5), similar to previous simulations \citep{diamessis2006,aghsaee2012,sakai2020}. 

From these simulations, it remains unclear if the continuous perturbations at $Re_{ISW}=900$, are indeed absolute/global instability, that follow the ISW or are continually generated convective instabilities. In order to gain some insight into the convective vs. absolute nature, we theoretically analyzed the local stability properties of the separated BBL under the ISW. \cite{diwan2012} have shown locally parallel stability theory is adequate to represent the primary linear regime of a separated laminar BBL. Here, our objective was to determine if an arbitrary localized disturbance, in the reference frame of the ISW, would propagate away from the generation site (convective) or if growth would occur where it was introduced (absolute). We computed the impulse-response function $G(x_{\ell},t)$ for selected profiles at different locations ($x_{\ell}$) along the streamwise axis by following \cite{alam2000}, which defines $G(x_{\ell},t)$ as the linear superposition of the discrete spectrum of $J$ unstable modes:
\begin{equation}
    G(x_{\ell},t)=\sum_{j=1}^{J}e^{i\left ( \alpha_j x_{\ell} - \omega_j t\right )}
    \label{eq:Gxt}
\end{equation}

Here, the complex frequency $\omega_j$ for each real $\alpha_j$ was computed from the Orr-Sommerfeld equation at each streamwise profile:
\begin{equation}
    \frac{i}{Re}\left ( \frac{\mathrm{d}^4 \hat{v}}{\mathrm{d} z^4 } - 2\alpha^2\frac{\mathrm{d}^2 \hat{v}}{\mathrm{d} z^2 } + \alpha^4 \hat{v}   \right )- \left (\alpha U(z) -\omega\right )\left (  \frac{\mathrm{d}^2 \hat{v}}{\mathrm{d} z^2 } -\alpha^2 \hat{v} \right ) - \alpha \frac{\mathrm{d}^2 U(z)}{\mathrm{d} z^2 }\hat{v}=0,
    \label{eq:OS1}
\end{equation}

with boundary conditions 

\begin{equation}
    \hat{v}(0)=\frac{d\hat{v}}{dz}(0)=0, \quad \hat{v}(z\rightarrow\infty)\rightarrow 0 , \quad \frac{d\hat{v}}{dz}(z\rightarrow\infty)\rightarrow 0 
    \label{eq:OS2}
\end{equation}

\noindent where $U(z)$ is the base velocity profile. Equation \ref{eq:OS1} was solved using a Chebyshev collocation method \citep{orszag1971}. Further details of the Orr-Sommerfeld equation solution and its validation are presented in Appendix \ref{appA}. The analysis was conducted in the frame of reference of the ISW, allowing the boundary-layer flow to be regarded as steady \citep{verschaeve2014}. This allowed us to invoke classical hydrodynamic stability theory \citep{drazin1981}. Thus, from the perspective of the rightward-propagating ISWs, the DNS velocity field was shifted by $-c$ for the stability analysis. We selected the base flow field $U(z)$ at a time immediately before any signs of instability were first observed. The results did not change when we repeated the analysis for different simulation times before the onset of instability (not shown). The analysis was carried out for $Re_{ISW}$ 300 and 900 (cases B and C, see Table \ref{table:1}). 

\begin{figure}
\centerline{\includegraphics[width=\textwidth]{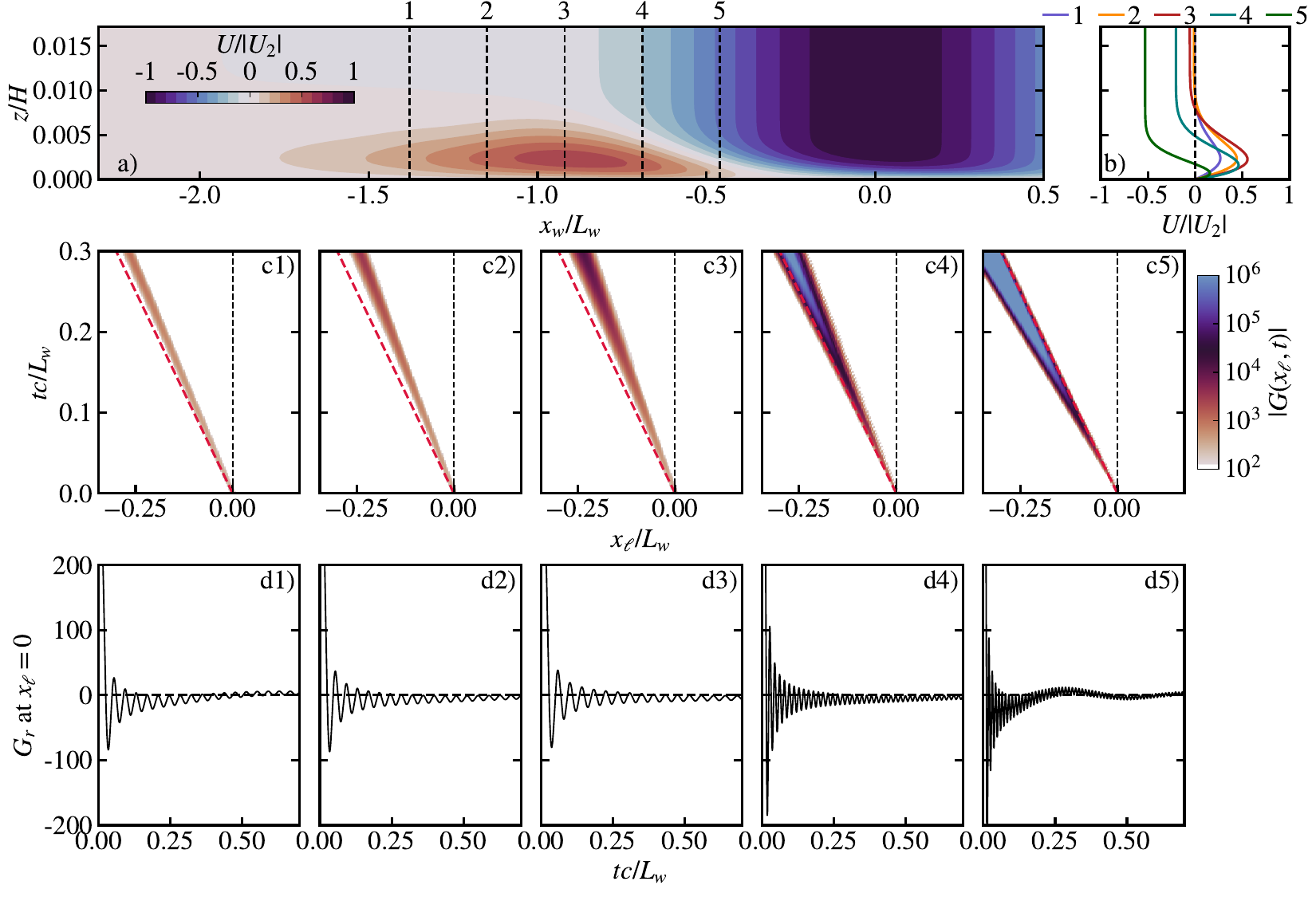}}
\caption{(a) Base flow ($U/\left | U_2 \right |$) and (b) selected profiles for the linear stability analysis. Base flow is shown in the wave reference frame, with $x_w=0$ corresponding to the ISW trough. (c) Amplitude of the impulse-response function $\left | G(x_{\ell},t) \right |$ at the selected locations $x_{\ell}$. The black and red dashed lines indicate respectively the path of a point moving with the ISW and a stationary point in the fixed bottom reference frame as seen in the ISW reference frame. (d) Real part of the impulse-response ($G_r$) function at $x_{\ell}/L_w=0$. Case C4, $Re_{ISW}=900$.}
\label{fig:imp-resp_Re3}
\end{figure}

\begin{figure}
\centerline{\includegraphics[width=\textwidth]{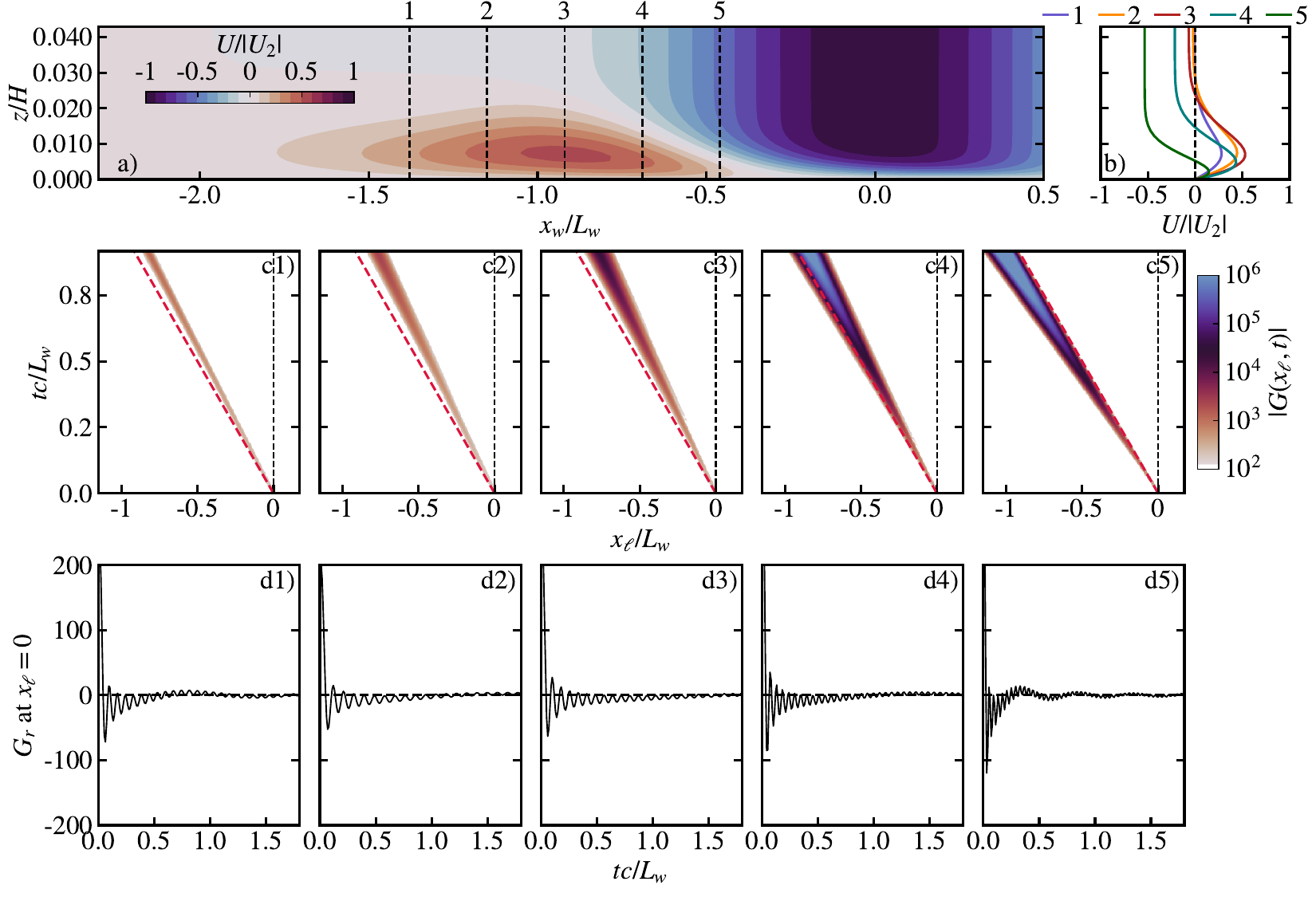}}
\caption{Same as Figure \ref{fig:imp-resp_Re3} for case B4, $Re_{ISW}=300$.}
\label{fig:imp-resp_Re2}
\end{figure}

The stability analysis predicted an impulse-response function with an unstable wavepacket growing while moving upstream of the ISW in the moving reference frame (i.e., falling behind the ISW) (Figure \ref{fig:imp-resp_Re3}c, Figure \ref{fig:imp-resp_Re2}c). The rapid decay of the impulse at $x_{\ell}/L_w=0$ (Figure \ref{fig:imp-resp_Re3}d, Figure \ref{fig:imp-resp_Re2}d), suggested that the flow was convectively unstable at both $Re_{ISW}$. Essentially, the group velocity of the linear instability is always lower than the ISW celerity. The linear stability analysis is in agreement with our $Re_{ISW}=300$ simulations and with the early stages of $Re_{ISW}=900$ simulations until $t/\tau\approx0.2$, as both showed instability-generated wave packets growing nearly in place in the fixed frame of reference, whilst falling behind the ISW (Figure \ref{fig:instRe2-ind}, Figure \ref{fig:instRe3-ind}); also in agreement with the noise-amplifier description suggested by \cite{verschaeve2014}.

Therefore, it is not clear why after $t/\tau\approx0.2$ a secondary instability tracked with the ISW in the numerical simulations at $Re_{ISW}=900$. This seeming discrepancy with the linear stability analysis raises the question: is it possible that at a large enough $Re_{ISW}$ (i.e., large enough amplification rate) the \emph{convective} instability can continuously amplify the background noise up to a finite amplitude (triggering vortex-shedding) within a distance $\sim O(L_w)$ from the ISW trough, thus seeming to track with the ISW? And, if that is the case, how can we differentiate such a mechanism from a self-sustained global mode whose main signature would also be a continuous emanation of instabilities amplifying into vortex shedding and trailing the ISW? These questions are beyond the scope of this work, but point to an alternative interpretation of BBL instability that challenges the global instability paradigm currently accepted \citep{boegman2019}. We consider that further research is necessary to unequivocally determine the nature of the instability at large $Re_{ISW}$. 

An argument in support of convective instability at $Re_{ISW}=900$ is found by analyzing the concept of a moving separated BBL in comparison to previous studies considering a steady laminar separation bubble. Local linear analyses of shear-layer profiles in the presence of a wall, representative of steady laminar separation bubbles, agree on the value of the relative reverse flow velocity $u_{rev}/U_{\infty}\approx12-25$\% for the onset of absolute instability \citep{hammond1998,alam2000,rodriguez2013}. We observe that for all our cases, even at $Re_{ISW}=$90 (not shown), the separated BBL exceeds this criterion (e.g., see profile 5 in Fig. \ref{fig:imp-resp_Re2}, $U_{\infty}=-0.53\left | U_2 \right |$, $u_{rev}=0.15\left | U_2 \right |$), despite reliably no absolute instability was observed at $Re_{ISW}=90$ (laminar) and $Re_{ISW}=300$ (convectively unstable). Therefore, using this $u_{rev}/U_{\infty}$ criterion is insufficient to determine if the instability is absolute. 

An important distinction here is the propagation of the separated BBL trailing the ISW which imposes unsteadiness on the flow. We hypothesize that as the separated shear layer was moving with the ISW, the rate of advection increased relative to the rate of instability growth, and so the instability was left behind before it could grow in place; at least during the primary linear stage. As $Re_{ISW}$ increased, the boundary layer becomes thinner (compare the vertical scales of Figure \ref{fig:imp-resp_Re3} and Figure \ref{fig:imp-resp_Re2}, panels a and b), increasing the shear and the total amplification of the potentially convective instability under the ISW. From this, we hypothesize that at a large enough $Re_{ISW}$ continuous instabilities can grow convectively up to a finite amplitude and trigger vortex shedding within the domain of the ISW without detaching from the ISW trough, mimicking a continuous self-sustained global mechanism. 

An example supporting the relevance of the moving separated BBL can be found in the case of an ISW propagating into an opposing current. \cite{sakai2020} identified the global mode oscillator from high-resolution 3D Large Eddy Simulation of an ISW of depression with very similar stability parameters to our $Re_{ISW}=300$ cases ($Re_{ISW}\approx310$, $P_{ISW}\approx0.1$, see Fig.~\ref{fig:paramspace}). They reported simulations, comparing the stability with and without the background current, and observed a convective (global) instability without (with) the current. \cite{stastna2008} found similar effects of an opposing background current on an ISW of elevation. The presence of the background current, for the BBL stability problem, has also been shown by \cite{becherer2020} to influence the location of the instability under the ISW, depending on the direction of the background current and the ISW polarity (elevation or depression). We suggest that an opposing background current can balance the advection of the instability, relative to the ISW, by delaying the ISW and causing it to propagate slower relative to the instability, giving more time for the instability to grow in place before being advected. 

Moreover, perhaps a more relevant effect is that the barotropic current adds an additional boundary layer (i.e., a laminar Blasius boundary layer, \cite{sakai2020}), which when superposed on that under the ISW could potentially favor a global mode excitation in comparison with the BBL induced by the ISW alone. \cite{sakai2016} indicate that the background current forces the separated BBL to reattach and develop a separation bubble under the ISW. Conversely, without a background current, the BBL remains separated at the wake of the ISW, as we observed in our $Re_{ISW}$ 90, 300 (Fig. \ref{fig:instRe2-ind}) and 900 (Fig. \ref{fig:noise}) simulations. Therefore, we consider that the propagation of an ISW against a background barotropic current poses a different boundary layer stability problem from that considered herein, which is expected to have different stability properties. 

As the Orr-Sommerfeld stability analysis above is linear, it can only describe the primary stage of the instability, and although our 2D simulations can capture the nonlinear evolution of the instability, they do not reproduce secondary 3D instabilities and the associated transition to turbulence. This precludes, for example, the possibility of secondary spanwise instability that may be convective or absolute \citep[e.g.][]{huerre1988,embacher2014}. It also precludes observing three-dimensional centrifugal instabilities expected to become unstable near $u_{rev}/U_{\infty}\approx7$\% \citep{rodriguez2013}, depending on the geometry of the separated BBL. Because of the wide variety of instability routes possible to transition to turbulence in laminar separation bubbles \citep[e.g.][]{rist2002,theofilis2011,embacher2014} further work is necessary to understand their relevance for the particular case of ISWs, with a moving separated BBL. Regardless of the nature of secondary instability processes that could ensue, our results point to the relevance of the moving separated BBL to favor convective primary instability within the parameter space evaluated herein. 

\subsection{Critical $Re_{ISW}$ for vortex shedding: numerical versus experimental}
The convective nature of the instability at $Re_{ISW}\approx300$, typical of lab-scale studies, might offer an explanation for discrepancies between the thresholds for vortex-shedding in numerical and laboratory experiments. The lab experiments by \cite{zahedi2021}, \cite{carr2006}, and \cite{carr2008} apparently have a common critical $Re_{ISW}\approx200$ for vortex shedding. This is much smaller than the threshold proposed from 2D numerical simulations by \cite{aghsaee2012} and \cite{diamessis2006}, also dependent on $P_{ISW}$ (Fig. \ref{fig:paramspace}). Recent DNS simulations by \cite{ellevold2023} showed good agreement with the lab experiments of \cite{carr2008} to predict a critical threshold for instability, disagreeing with the threshold proposed by \cite{aghsaee2012}. Given the noise amplifier behavior and the different background seed noise levels in the lab and different numerical solvers, these differences could be expected to influence the timing for instability growth to finite observable levels. Numerical solvers often have lower background noise levels than in the lab, hence larger amplification (i.e., larger $Re_{ISW}$) would be necessary in the numerical domains before instabilities are visually observable. For example, \cite{verschaeve2014} estimated the numerical noise in \cite{aghsaee2012} to be more than two orders of magnitude lower than in the experiments of \cite{carr2008}. Also, different numerical solvers have different background noise levels depending on the truncation error of the numerical approximations. A potential approach to evaluate this hypothesis would be by conducting parallel numerical and lab experiments, both at comparable and sufficiently large $Re$, with a controlled similar and constant level of background noise, such that instability can be characterized in both cases under similar conditions. This can be complemented with numerical simulations of a given ISW propagating through different levels of constant and uniform background noise, such that the effect of noise amplitude on the BBL stability can be investigated. 

The discussion above neglects other possible processes relevant to boundary layer stability, like wall roughness \citep{carr2010A,harnanan2017}, which might naturally be present in the lab but is not in the present simulations. For example, wall roughness might provide a mechanism for introducing seeding perturbations in the BBL susceptible to being convectively amplified beneath the ISW. Future research will consider the effects of bottom roughness on BBL instability under ISWs. 

\subsection{Implications for sediment resuspension}
The initial motivation for this work was to understand the nature of the instability as relates to the potential for sediment resuspension and transport \citep{aghsaee2015,zulberti2020}. Sediment resuspension induced by the periodic amplification of the $Re_{ISW}=300$ instability, falling behind the ISW, would be expected to be very different from that triggered by the continuous vortex shedding trailing the ISW at $Re_{ISW}=900$.

As shown in Fig. \ref{fig:noise}, instability amplification and vortex shedding were accompanied by an increase in the instantaneous bed stress, with the potential to resuspend sediments. Free-stream perturbations and $Re_{ISW}$ can influence the location where the instability reaches a finite amplitude and triggers vortex shedding under the ISW and so, they can also be expected to influence the location of maximum sediment resuspension under the ISW. Bringing these new ingredients into consideration might help to interpret field observations which often show maximum sediment resuspension trailing the ISW \cite[e.g.][]{johnson2001,bogucki2005,becherer2020} and in other cases show it beneath the wave trough \citep[e.g.][]{quaresma2007,zulberti2020}.

The most recent field measurements by \cite{zulberti2020} were more detailed near the bed, and so are considered first. They described their observations in terms of a pumping mechanism resulting from the alternating compression and expansion of the BBL forced by trains of ISWs. The waves resuspended sediment during the compression phase under the front shoulder of the ISW and then pumped it upwards into the water column during the expansion phase under the rear shoulder. The maximum near-bed sediment concentrations were observed under the ISW trough. They did not identify flow separation, nor global instability mechanisms and their observations occurred at a comparatively high $Re_{ISW}\approx15000$. The ISWs propagated through a highly turbulent boundary layer, energetic enough to sustain an inertial sub-layer \citep{zulberti2022}. A key feature of their observations was that no separation bubble, and thus, no shear instability was observed. We believe it is likely that the combined effect of large $Re$ and free-stream turbulence (highly turbulent BBL) was large enough to dwarf, and potentially suppress the separation bubble \citep[e.g.][]{balzer2016,simoni2017}. In support of this, an example of both effects can be seen in Fig. \ref{fig:noise}, where the reverse-flow vortex due to the separated BBL at $Re_{ISW}=900$ is dwarfed by the vortex-shedding closer to the ISW trough due to the increased level of wake perturbations (panel c) and by increasing $Re_{ISW}$ (panel d). 

On the other hand, field observations reporting sediment resuspension trailing an ISW of depression \citep{johnson2001,bogucki2005,becherer2020} tend to have more modest $Re_{ISW}\sim O(2500-3000)$ than in \cite{zulberti2020}. Also, these sites can be assumed to have much thinner and less energetic turbulent boundary layers, as significant sediment was not observed to be in suspension prior to the passage of the ISW. Therefore, it is also reasonable to assume lower levels of external free-stream turbulence, which, along with lower $Re_{ISW}$ might favor instability breaking into vortex shedding behind the ISW trough. In comparison to lab settings, these cases have much larger $Re_{ISW}$, such that continuously trailing vortex shedding might be more likely to be observed, as long as it is not significantly influenced by other environmental factors, like free-stream turbulence or a barotropic current. Unfortunately, none of these studies include detailed measurements of the near-bed turbulent field and so we cannot make any further conclusions. 

\subsection{Relaminarization}

We have shown the energy of the instability to become reduced under the front shoulder of the ISW as work is done against the longitudinal stretching of the base flow for $Re_{ISW}=300$ (Figure \ref{fig:instRe2-train}). However, such behavior is different from what was observed in the field by \cite{zulberti2020}, where turbulent kinetic energy and shear production increased over two orders of magnitude under the front shoulder, reaching a maximum under the wave trough (their Figure 3). In a self-similar accelerating boundary layer, relaminarization can be expected if the acceleration parameter $K=(\nu/U^2_{\infty})(dU_{\infty}/dx)\geq3\times10^{-6}$ \citep{narasimha1979}. In terms of readily available field parameters for ISWs reported by \cite{zulberti2020}, the parameter $K$ would be $K\sim\nu/(U_{2}L_w)$, which can be estimated as $K\approx3\times10^{-9}$. This is three orders of magnitude lower than the critical value, justifying no relaminarization in their observations. Conversely, the ISWs simulated here at $Re_{ISW}=300$ have an associated $K\approx1\times10^{-5}$, which is large enough to expect relaminarization, in agreement with our simulations showing an analogous $E_v$ reduction under the ISW front shoulder. This suggests relaminarization under the front shoulder is an additional feature that might be different between field and lab-scale ISWs. 

\section{Conclusions}
This research investigated the effect of the Reynolds number and ISW periodicity on the stability properties of the BBL under ISWs. The boundary layer stability showed a strong dependence on $Re_{ISW}$, which determined the stability regime and rate of vortex shedding in the BBL. The effect of wave periodicity was more subtle, only significant around a regime captured by $Re_{ISW}=300$, where the convective instability triggered by an individual ISW was mild enough to not trigger vortex-shedding before the next ISWs in the train periodically reinforced it. 

Numerical simulations predicted laminar and convectively unstable regimes at $Re_{ISW}$ 90 and 300 respectively. For higher $Re_{ISW}=900$, the BBL was initially convectively unstable, and then instabilities continuously emanated from the separated BBL tracking with the ISW, which is typically associated with a global instability. The initial convective instability is in agreement with local linear stability theory at both $Re_{ISW}$ 300 and 900, which essentially predicts that the instability group speed is always lower than the ISW phase speed. We hypothesized that continuous convective amplification was mimicking a global mechanism at $Re_{ISW}$ 900. Further research is necessary to unequivocally determine the nature of the instability at higher $Re_{ISW}$. 

The convective instability at typical lab-scale $Re_{ISW}\approx300$ shows a noise-amplifier behavior of the flow, which offers an explanation for discrepancies in the critical threshold for vortex shedding between lab and different numerical simulations due to differences in the background noise.

Our simulations show that increasing levels of free-stream perturbations and larger $Re_{ISW}$ shift the location of vortex shedding (and enhanced bed shear stress) closer to the ISW trough, with potential consequences for the location of the maximum sediment resuspension under the ISW. 

From our simulations, we illustrate additional ingredients that might influence BBL stability under ISWs, including (i) free-stream perturbations, (ii) Reynolds number effects, (iii) background barotropic currents, and (iv) relaminarization. These might be particularly relevant to further compare against other numerical, experimental, and field observations. \\ \\

\textbf{Acknowledgements}\\
The authors thank Marek Stastna for discussions. This research made use of the high-performance computing clusters of Compute Ontario (computeontario.ca) and the Digital Research Alliance of Canada (alliancecan.ca). 

\textbf{Funding:} The research was funded by NSERC Discovery Grants to L.B. and by Queen’s University.

\textbf{Competing interests:} The authors report no conflict of interest.


\appendix
\section{Orr-Sommerfeld solver}\label{appA}
The Orr-Sommerfeld equation \ref{eq:OS1} represents a generalized eigenvalue problem in matrix form
\begin{equation}
    \mathsfbi{A}\mathbf{\hat{v}}=\omega \mathsfbi{B}\mathbf{\hat{v}}
    \label{eq:A1}
\end{equation}
with $\mathbf{\hat{v}}$ as the eigenvector and the complex frequency $\omega$ as the eigenvalue. Equation \ref{eq:A1} was solved using a Chebyshev collocation method on 250 nodes, following \cite{orszag1971}. Derivatives were computed using Chebyshev differentiation matrices following \cite{weideman2000}. The code solves the temporal eigenvalue problem, returning all the sets of modes associated with a given real wavenumber $\alpha$, from which we selected $\omega$ for the most unstable eigenmode (largest $\Im({\omega})$). The DNS-simulated near-bed velocity profile $U(z)$ and the grid used for the stability analysis were extended further away from the wall, so the velocity profile smoothly increased to free-stream conditions. 

We validated the code by comparing the most unstable eigenvalue for the Blasius boundary layer to that reported by \cite{gaster1978}. We found agreement with their results over the range $Re_{\delta^*}$ 500-3000 to the 6th digit for the real and imaginary parts. 

We also validated our implementation by solving the impulse-response function for the reverse-flow profiles analyzed by \cite{alam1997} and \cite{alam2000}:
\begin{equation}
    \frac{u}{U_{\infty}}=\tanh(z) - 2A\frac{\tanh(z/B)}{\cosh^2(z/B)}
    \label{eq:A2}
\end{equation}
where constants $A$ and $B$ control the amount of reverse flow and the distance of the inflection point from the wall, respectively. We compared our results with those given in figures 21-24 in \cite{alam2000} and figures 4-6 in \cite{alam1997}, with quite good agreement in all cases.

\bibliographystyle{jfm}
\bibliography{Biblio/main}

\begin{thebibliography}{51}
\expandafter\ifx\csname natexlab\endcsname\relax\def\natexlab#1{#1}\fi
\def\au#1{#1} \def\ed#1{#1} \def\yr#1{#1}\def\at#1{#1}\def\jt#1{\textit{#1}}
  \def\bt#1{#1}\def\bvol#1{\textbf{#1}} \def\vol#1{#1} \def\pg#1{#1}
  \def\publ#1{#1}\def\arxiv#1{#1}\def\org#1{#1}\def\st#1{\textit{#1}}

\bibitem[Aghsaee \& Boegman(2015)]{aghsaee2015}
{\sc \au{Aghsaee, P.} \& \au{Boegman, L.}} \yr{2015}  \at{Experimental
  investigation of sediment resuspension beneath internal solitary waves of
  depression: {Solitary} wave-induced resuspension}.  \jt{Journal of
  Geophysical Research: Oceans}  \bvol{120}~(5),  \pg{3301--3314}.

\bibitem[Aghsaee {\em et~al.\/}(2012)Aghsaee, Boegman, Diamessis \&
  Lamb]{aghsaee2012}
{\sc \au{Aghsaee, P.}, \au{Boegman, L.}, \au{Diamessis, P.~J.} \& \au{Lamb,
  K.~G.}} \yr{2012}  \at{Boundary-layer-separation-driven vortex shedding
  beneath internal solitary waves of depression}.  \jt{Journal of Fluid
  Mechanics}  \bvol{690},  \pg{321--344}.

\bibitem[Alam \& Sandham(1997)]{alam1997}
{\sc \au{Alam, M} \& \au{Sandham, ND}} \yr{1997} Simulation of laminar
  separation bubble instabilities.  \bt{In {\em Direct and Large-Eddy
  Simulation II: Proceedings of the ERCOFTAC Workshop held in Grenoble, France,
  16--19 September 1996\/}},  \pg{pp. 125--136}. Springer.

\bibitem[Alam \& Sandham(2000)]{alam2000}
{\sc \au{Alam, M.} \& \au{Sandham, N.~D.}} \yr{2000}  \at{Direct numerical
  simulation of ‘short’laminar separation bubbles with turbulent
  reattachment}.  \jt{Journal of Fluid Mechanics}  \bvol{410},  \pg{1--28}.

\bibitem[Balzer \& Fasel(2016)]{balzer2016}
{\sc \au{Balzer, W.} \& \au{Fasel, H.~F.}} \yr{2016}  \at{Numerical
  investigation of the role of free-stream turbulence in boundary-layer
  separation}.  \jt{Journal of Fluid Mechanics}  \bvol{801},  \pg{289--321}.

\bibitem[Becherer {\em et~al.\/}(2020)Becherer, Moum, Colosi, Lerczak \&
  McSweeney]{becherer2020}
{\sc \au{Becherer, J.}, \au{Moum, J.~N.}, \au{Colosi, J.~A.}, \au{Lerczak,
  J.~A.} \& \au{McSweeney, J.~M.}} \yr{2020}  \at{Turbulence asymmetries in
  bottom boundary layer velocity pulses associated with onshore-propagating
  nonlinear internal waves}.  \jt{Journal of Physical Oceanography}
  \bvol{50}~(8),  \pg{2373--2391}.

\bibitem[Boegman \& Stastna(2019)]{boegman2019}
{\sc \au{Boegman, L.} \& \au{Stastna, M.}} \yr{2019}  \at{Sediment
  {Resuspension} and {Transport} by {Internal} {Solitary} {Waves}}.  \jt{Annual
  Review of Fluid Mechanics}  \bvol{51}~(1),  \pg{129--154}.

\bibitem[Bogucki \& Redekopp(1999)]{bogucki1999}
{\sc \au{Bogucki, D.~J.} \& \au{Redekopp, L.~G.}} \yr{1999}  \at{A mechanism
  for sediment resuspension by internal solitary waves}.  \jt{Geophysical
  Research Letters}  \bvol{26}~(9),  \pg{1317--1320}.

\bibitem[Bogucki {\em et~al.\/}(2005)Bogucki, Redekopp \& Barth]{bogucki2005}
{\sc \au{Bogucki, D.~J.}, \au{Redekopp, L.~G.} \& \au{Barth, J.}} \yr{2005}
  \at{Internal solitary waves in the coastal mixing and optics 1996 experiment:
  Multimodal structure and resuspension}.  \jt{Journal of Geophysical Research:
  Oceans}  \bvol{110}~(C2).

\bibitem[Carr \& Davies(2006)]{carr2006}
{\sc \au{Carr, M.} \& \au{Davies, P.~A.}} \yr{2006}  \at{The motion of an
  internal solitary wave of depression over a fixed bottom boundary in a
  shallow, two-layer fluid}.  \jt{Physics of Fluids}  \bvol{18}~(1),
  \pg{016601}.

\bibitem[Carr {\em et~al.\/}(2008)Carr, Davies \& Shivaram]{carr2008}
{\sc \au{Carr, M.}, \au{Davies, P.~A.} \& \au{Shivaram, P.}} \yr{2008}
  \at{Experimental evidence of internal solitary wave-induced global
  instability in shallow water benthic boundary layers}.  \jt{Physics of
  Fluids}  \bvol{20}~(6),  \pg{066603}.

\bibitem[Carr {\em et~al.\/}(2010)Carr, Stastna \& Davies]{carr2010A}
{\sc \au{Carr, M.}, \au{Stastna, M.} \& \au{Davies, P.~A.}} \yr{2010}
  \at{Internal solitary wave-induced flow over a corrugated bed}.  \jt{Ocean
  dynamics}  \bvol{60},  \pg{1007--1025}.

\bibitem[Chomaz(2005)]{chomaz2005}
{\sc \au{Chomaz, J.}} \yr{2005}  \at{Global instabilities in spatially
  developing flows: Non-normality and nonlinearity}.  \jt{Annual Review of
  Fluid Mechanics}  \bvol{37}~(1),  \pg{357--392}.

\bibitem[Deepwell {\em et~al.\/}(2021)Deepwell, Clarry, Subich \&
  Stastna]{deepwell2021}
{\sc \au{Deepwell, D.}, \au{Clarry, C.}, \au{Subich, C.} \& \au{Stastna, M.}}
  \yr{2021}  \at{Vortex generation due to internal solitary wave propagation
  past a sidewall constriction}.  \jt{Journal of Fluid Mechanics}  \bvol{913},
  \pg{A47--26}.

\bibitem[Diamessis \& Redekopp(2006)]{diamessis2006}
{\sc \au{Diamessis, P.~J.} \& \au{Redekopp, L.~G.}} \yr{2006}  \at{Numerical
  {Investigation} of {Solitary} {Internal} {Wave}-{Induced} {Global}
  {Instability} in {Shallow} {Water} {Benthic} {Boundary} {Layers}}.
  \jt{Journal of Physical Oceanography}  \bvol{36}~(5),  \pg{784--812}.

\bibitem[Diwan \& Ramesh(2012)]{diwan2012}
{\sc \au{Diwan, S.~S.} \& \au{Ramesh, O.~N.}} \yr{2012}  \at{Relevance of local
  parallel theory to the linear stability of laminar separation bubbles}.
  \jt{Journal of fluid mechanics}  \bvol{698},  \pg{468--478}.

\bibitem[Drazin \& Reid(1981)]{drazin1981}
{\sc \au{Drazin, P.~G.} \& \au{Reid, W.~H.}} \yr{1981} {\em Hydrodynamic
  {Stability}\/}, 2nd edn.  \publ{Cambridge University Press}.

\bibitem[Dunphy {\em et~al.\/}(2011)Dunphy, Subich \& Stastna]{dunphy2011}
{\sc \au{Dunphy, M.}, \au{Subich, C.} \& \au{Stastna, M.}} \yr{2011}
  \at{Spectral methods for internal waves: indistinguishable density profiles
  and double-humped solitary waves}.  \jt{Nonlinear Processes in Geophysics}
  \bvol{18}~(3),  \pg{351--358}.

\bibitem[Ellevold \& Grue(2023)]{ellevold2023}
{\sc \au{Ellevold, T.~J.} \& \au{Grue, J.}} \yr{2023}  \at{Calculation of
  internal-wave-driven instability and vortex shedding along a flat bottom}.
  \jt{Journal of Fluid Mechanics}  \bvol{966},  \pg{A40}.

\bibitem[Embacher \& Fasel(2014)]{embacher2014}
{\sc \au{Embacher, M.} \& \au{Fasel, H.~F.}} \yr{2014}  \at{Direct numerical
  simulations of laminar separation bubbles: investigation of absolute
  instability and active flow control of transition to turbulence}.
  \jt{Journal of fluid mechanics}  \bvol{747},  \pg{141--185}.

\bibitem[Gaster(1978)]{gaster1978}
{\sc \au{Gaster, M}} \yr{1978}  \at{Series representation of the eigenvalues of
  the orr-sommerfeld equation}.  \jt{Journal of Computational Physics}
  \bvol{29}~(2),  \pg{147--162}.

\bibitem[Ghassemi {\em et~al.\/}(2022)Ghassemi, Zahedi \&
  Boegman]{ghassemi2022}
{\sc \au{Ghassemi, A.}, \au{Zahedi, S.} \& \au{Boegman, L.}} \yr{2022}
  \at{Bolus formation from fission of nonlinear internal waves over a mild
  slope}.  \jt{Journal of Fluid Mechanics}  \bvol{932},  \pg{A50}.

\bibitem[Hammond \& Redekopp(1998)]{hammond1998}
{\sc \au{Hammond, D.~A.} \& \au{Redekopp, L.~G.}} \yr{1998}  \at{Local and
  global instability properties of separation bubbles}.  \jt{European Journal
  of Mechanics - B/Fluids}  \bvol{17}~(2),  \pg{145--164}.

\bibitem[Harnanan {\em et~al.\/}(2017)Harnanan, Stastna \&
  Soontiens]{harnanan2017}
{\sc \au{Harnanan, S.}, \au{Stastna, M.} \& \au{Soontiens, N.}} \yr{2017}
  \at{The effects of near-bottom stratification on internal wave induced
  instabilities in the boundary layer}.  \jt{Physics of Fluids}  \bvol{29}~(1),
   \pg{016602}.

\bibitem[Hartharn-Evans {\em et~al.\/}(2022)Hartharn-Evans, Carr, Stastna \&
  Davies]{hartharn-evans2022A}
{\sc \au{Hartharn-Evans, S.~G.}, \au{Carr, M.}, \au{Stastna, M.} \& \au{Davies,
  P.~A.}} \yr{2022}  \at{Stratification effects on shoaling internal solitary
  waves}.  \jt{Journal of Fluid Mechanics}  \bvol{933},  \pg{A19}.

\bibitem[Helfrich \& Melville(2006)]{helfrich2006}
{\sc \au{Helfrich, K.~R.} \& \au{Melville, W.~K.}} \yr{2006}  \at{Long
  nonlinear internal waves}.  \jt{Annu. Rev. Fluid Mech.}  \bvol{38},
  \pg{395--425}.

\bibitem[Huerre(1988)]{huerre1988}
{\sc \au{Huerre, P.}} \yr{1988} On the absolute/convective nature of primary
  and secondary instabilities.  \bt{In {\em Propagation in Systems Far from
  Equilibrium\/} (ed. \ed{J.~E. Wesfreid, Helmut~R. Brand, P.~Manneville,
  G.~Albinet \& N.~Boccara})},  \pg{pp. 340--353}.  \publ{Berlin, Heidelberg:
  Springer Berlin Heidelberg}.

\bibitem[Huerre \& Monkewitz(1990)]{huerre1990}
{\sc \au{Huerre, P.} \& \au{Monkewitz, P.~A.}} \yr{1990}  \at{Local and
  {Global} {Instabilities} in {Spatially} {Developing} {Flows}}.  \jt{Annual
  Review of Fluid Mechanics}  \bvol{22}~(1),  \pg{473--537}.

\bibitem[Johnson {\em et~al.\/}(2001)Johnson, Weidemann \& Pegau]{johnson2001}
{\sc \au{Johnson, D.~R.}, \au{Weidemann, A.} \& \au{Pegau, W.~S.}} \yr{2001}
  \at{Internal tidal bores and bottom nepheloid layers}.  \jt{Continental Shelf
  Research}  \bvol{21}~(13-14),  \pg{1473--1484}.

\bibitem[Lamb(2014)]{lamb2014}
{\sc \au{Lamb, K.~G.}} \yr{2014}  \at{Internal wave breaking and dissipation
  mechanisms on the continental slope/shelf}.  \jt{Annual Review of Fluid
  Mechanics}  \bvol{46},  \pg{231--254}.

\bibitem[Michallet \& Ivey(1999)]{michallet1999}
{\sc \au{Michallet, H.} \& \au{Ivey, G.~N.}} \yr{1999}  \at{Experiments on
  mixing due to internal solitary waves breaking on uniform slopes}.
  \jt{Journal of Geophysical Research: Oceans}  \bvol{104}~(C6),
  \pg{13467--13477},  \arxiv{arXiv:
  https://agupubs.onlinelibrary.wiley.com/doi/pdf/10.1029/1999JC900037}.

\bibitem[Narasimha \& Sreenivasan(1979)]{narasimha1979}
{\sc \au{Narasimha, R.} \& \au{Sreenivasan, K.~R.}} \yr{1979}
  \at{Relaminarization of fluid flows}.  \jt{Advances in applied mechanics}
  \bvol{19},  \pg{221--309}.

\bibitem[Orszag(1971)]{orszag1971}
{\sc \au{Orszag, S~A}} \yr{1971}  \at{Accurate solution of the orr--sommerfeld
  stability equation}.  \jt{Journal of Fluid Mechanics}  \bvol{50}~(4),
  \pg{689--703}.

\bibitem[Quaresma {\em et~al.\/}(2007)Quaresma, Vitorino, Oliveira \&
  da~Silva]{quaresma2007}
{\sc \au{Quaresma, L.~S.}, \au{Vitorino, J.}, \au{Oliveira, A.} \&
  \au{da~Silva, J.}} \yr{2007}  \at{Evidence of sediment resuspension by
  nonlinear internal waves on the western {Portuguese} mid-shelf}.  \jt{Marine
  Geology}  \bvol{246}~(2-4),  \pg{123--143}.

\bibitem[Rist \& Maucher(2002)]{rist2002}
{\sc \au{Rist, U.} \& \au{Maucher, U.}} \yr{2002}  \at{Investigations of
  time-growing instabilities in laminar separation bubbles}.  \jt{European
  Journal of Mechanics-B/Fluids}  \bvol{21}~(5),  \pg{495--509}.

\bibitem[Rodr{\'\i}guez {\em et~al.\/}(2013)Rodr{\'\i}guez, Gennaro \&
  Juniper]{rodriguez2013}
{\sc \au{Rodr{\'\i}guez, D.}, \au{Gennaro, E.~M.} \& \au{Juniper, M.~P.}}
  \yr{2013}  \at{The two classes of primary modal instability in laminar
  separation bubbles}.  \jt{Journal of Fluid Mechanics}  \bvol{734},  \pg{R4}.

\bibitem[Sakai {\em et~al.\/}(2016)Sakai, Diamessis \& Jacobs]{sakai2016}
{\sc \au{Sakai, T.}, \au{Diamessis, P.~J.} \& \au{Jacobs, G.~B.}} \yr{2016}
  Large eddy simulations of turbulence under internal solitary waves of
  depression.  \bt{In {\em International Symposium on Stratified Flows\/}},
  \st{1},  \vol{vol.~1}. UC San Diego.

\bibitem[Sakai {\em et~al.\/}(2020)Sakai, Diamessis \& Jacobs]{sakai2020}
{\sc \au{Sakai, T.}, \au{Diamessis, P.~J.} \& \au{Jacobs, G.~B.}} \yr{2020}
  \at{Self-sustained instability, transition, and turbulence induced by a long
  separation bubble in the footprint of an internal solitary wave. {I}. {Flow}
  topology}.  \jt{Physical Review Fluids}  \bvol{5}~(10),  \pg{103801}.

\bibitem[Schmid \& Henningson(2001)]{schmid2001}
{\sc \au{Schmid, P.~J.} \& \au{Henningson, D.~S.}} \yr{2001} {\em Stability and
  {Transition} in {Shear} {Flows}\/},  \st{Applied {Mathematical} {Sciences}},
  \vol{vol. 142}.  \publ{New York, NY: Springer New York}.

\bibitem[Simoni {\em et~al.\/}(2017)Simoni, Lengani, Ubaldi, Zunino \&
  Dellacasagrande]{simoni2017}
{\sc \au{Simoni, D.}, \au{Lengani, D.}, \au{Ubaldi, M.}, \au{Zunino, P.} \&
  \au{Dellacasagrande, M.}} \yr{2017}  \at{Inspection of the dynamic properties
  of laminar separation bubbles: free-stream turbulence intensity effects for
  different {Reynolds} numbers}.  \jt{Experiments in Fluids}  \bvol{58}~(6),
  \pg{66}.

\bibitem[Stastna \& Lamb(2008)]{stastna2008}
{\sc \au{Stastna, M.} \& \au{Lamb, K.~G.}} \yr{2008}  \at{Sediment resuspension
  mechanisms associated with internal waves in coastal waters}.  \jt{Journal of
  Geophysical Research}  \bvol{113}~(C10),  \pg{C10016}.

\bibitem[Subich {\em et~al.\/}(2013)Subich, Lamb \& Stastna]{subich2013}
{\sc \au{Subich, C.~J.}, \au{Lamb, K.~G.} \& \au{Stastna, M.}} \yr{2013}
  \at{Simulation of the navier–stokes equations in three dimensions with a
  spectral collocation method}.  \jt{International Journal for Numerical
  Methods in Fluids}  \bvol{73}~(2),  \pg{103--129},  \arxiv{arXiv:
  https://onlinelibrary.wiley.com/doi/pdf/10.1002/fld.3788}.

\bibitem[Theofilis(2011)]{theofilis2011}
{\sc \au{Theofilis, V.}} \yr{2011}  \at{Global {Linear} {Instability}}.
  \jt{Annual Review of Fluid Mechanics}  \bvol{43}~(1),  \pg{319--352}.

\bibitem[Torrence \& Compo(1998)]{torrence1998}
{\sc \au{Torrence, C.} \& \au{Compo, G.~P.}} \yr{1998}  \at{A {Practical}
  {Guide} to {Wavelet} {Analysis}}.  \jt{Bulletin of the American
  Meteorological Society}  \bvol{79}~(1),  \pg{61--78}.

\bibitem[Trowbridge \& Lentz(2018)]{trowbridge2018}
{\sc \au{Trowbridge, J.~H.} \& \au{Lentz, S.~J.}} \yr{2018}  \at{The {Bottom}
  {Boundary} {Layer}}.  \jt{Annual Review of Marine Science}  \bvol{10}~(1),
  \pg{397--420}.

\bibitem[Turkington {\em et~al.\/}(1991)Turkington, Eydeland \&
  Wang]{turkington1991}
{\sc \au{Turkington, B.}, \au{Eydeland, A.} \& \au{Wang, S.}} \yr{1991}  \at{A
  {Computational} {Method} for {Solitary} {Internal} {Waves} in a
  {Continuously} {Stratified} {Fluid}}.  \jt{Studies in Applied Mathematics}
  \bvol{85}~(2),  \pg{93--127}.

\bibitem[Verschaeve \& Pedersen(2014)]{verschaeve2014}
{\sc \au{Verschaeve, J. C.~G.} \& \au{Pedersen, G.~K.}} \yr{2014}  \at{Linear
  stability of boundary layers under solitary waves}.  \jt{Journal of Fluid
  Mechanics}  \bvol{761},  \pg{62--104}.

\bibitem[Weideman \& Reddy(2000)]{weideman2000}
{\sc \au{Weideman, J.~A.} \& \au{Reddy, S.~C.}} \yr{2000}  \at{A matlab
  differentiation matrix suite}.  \jt{ACM Transactions on Mathematical Software
  (TOMS)}  \bvol{26}~(4),  \pg{465--519}.

\bibitem[Zahedi {\em et~al.\/}(2021)Zahedi, Aghsaee \& Boegman]{zahedi2021}
{\sc \au{Zahedi, S.}, \au{Aghsaee, P.} \& \au{Boegman, L.}} \yr{2021}
  \at{Internal solitary wave bottom boundary layer dissipation}.  \jt{Physical
  Review Fluids}  \bvol{6}~(7),  \pg{074802}.

\bibitem[Zulberti {\em et~al.\/}(2020)Zulberti, Jones \& Ivey]{zulberti2020}
{\sc \au{Zulberti, A.}, \au{Jones, N.~L.} \& \au{Ivey, G.~N.}} \yr{2020}
  \at{Observations of enhanced sediment transport by nonlinear internal waves}.
   \jt{Geophysical Research Letters}  \bvol{47}~(19),  \pg{e2020GL088499},
  e2020GL088499 2020GL088499,  \arxiv{arXiv:
  https://agupubs.onlinelibrary.wiley.com/doi/pdf/10.1029/2020GL088499}.

\bibitem[Zulberti {\em et~al.\/}(2022)Zulberti, Jones, Rayson \&
  Ivey]{zulberti2022}
{\sc \au{Zulberti, A.~P.}, \au{Jones, N.~L.}, \au{Rayson, M.~D.} \& \au{Ivey,
  G.~N.}} \yr{2022}  \at{Mean and turbulent characteristics of a bottom
  mixing-layer forced by a strong surface tide and large amplitude internal
  waves}.  \jt{Journal of Geophysical Research: Oceans}  \bvol{127}~(1),
  \pg{e2020JC017055}.

\end{thebibliography}

\end{document}